\documentclass[
preprint,
 amsmath,amssymb,
 aps,
 prl,
 superscriptaddress,
]{revtex4-2}

\usepackage{float}
\usepackage{dsfont}
\usepackage{lipsum}
\usepackage{xcolor}
\usepackage{enumitem}
\usepackage{notes2bib}

\setlist[itemize]{noitemsep, nolistsep}
\usepackage{graphicx}
\usepackage{dcolumn}
\usepackage{bm}
\usepackage{hyperref}

\begin{document}

\title{Optimal Rejection-Free Path Sampling}

\author{Gianmarco Lazzeri}
\email{lazzeri@fias.uni-frankfurt.de}
\affiliation{Institute of Biochemistry, Goethe University Frankfurt}
\affiliation{Frankfurt Institute for Advanced Studies}

\author{Peter G. Bolhuis}
\email{P.G.Bolhuis@uva.nl}
\affiliation{Van 't Hoff Institute for Molecular Sciences, University of Amsterdam}

\author{Roberto Covino}
\email{covino@fias.uni-frankfurt.de}
\affiliation{Institute of Computer Science, Goethe University Frankfurt}
\affiliation{Frankfurt Institute for Advanced Studies}

\date{\today}

\begin{abstract}

We propose an efficient novel path sampling-based framework designed to accelerate the investigation of rare events in complex molecular systems. A key innovation is the shift from sampling restricted path ensemble distributions, as in transition path sampling, to directly sampling the distribution of shooting points. This allows for a rejection-free algorithm that samples the entire path ensemble efficiently. 
Optimal sampling is achieved by applying a selection bias that is the inverse of the free energy along a reaction coordinate. The optimal reaction coordinate, the committor, is iteratively constructed as a neural network using AI for Molecular Mechanism Discovery (AIMMD), concurrently with the free energy profile, which is obtained through reweighting the sampled path ensembles. 
We showcase our algorithm on theoretical and molecular bechnmarks, and demonstrate how it provides at the same time molecular mechanism, free energy, and rates at a moderate computational cost. 
\end{abstract}

\maketitle


\noindent{\textit{Introduction}} --- Many critical molecular processes involve spontaneous structural rearrangements, modeled as infrequent stochastic transitions in a complex energy landscape \cite{peters2017reaction}. Understanding the mechanisms, thermodynamics, and kinetics of such rare transitions is essential for, e.g., elucidating  functional mechanisms of vital biomolecular complexes \cite{whitford2012biomolecular} or providing a foundation for designing novel enzymes \cite{de2016role}.

\begin{figure}
\includegraphics[width=0.45\textwidth]{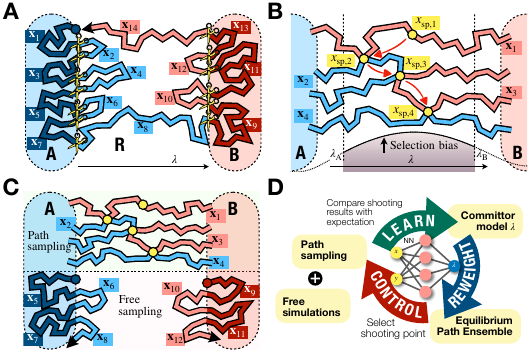}
\caption{\label{fig:1} Theory.
\textbf{(A)} The PE sampled from an indefinitely long equilibrium trajectory, split in internal segments (blue: in-$\mathrm{A}$, red: in-$\mathrm{B}$) and excursions (light blue: from $\mathrm{A}$, pink: from $\mathrm{B}$) at the boundary of the metastable states $\mathrm{A}$ and $\mathrm{B}$. Most excursions are short. 
$\textbf{x}_8$ and $\textbf{x}_{14}$ are transitions. 
\textbf{(B)} RFPS as a Markov chain of SPs (gold dots connected by red arrows). Paths are the (by)products of SP generation moves and sample the PE upon reweighting. SPs are selected from previous paths inversely proportional to the Boltzmann distribution projected on $\lambda$ between $\lambda_{\mathrm{A}}$ and $\lambda_{\mathrm{B}}$. \textbf{(C)} AIMMD enhanced sampling as the combination of RFPS in the reactive region (for generating large excursions and transitions) and ``free'' simulations around the metastable states (for generating internal segments and most small excursions). The simulations run in parallel. \textbf{(D)} The RFPS-AIMMD cycle iteratively optimizes the committor model, the PE estimate, the Boltzmann distribution, and the sampling.
}
\end{figure}

Although molecular dynamics (MD) is an attractive way of investigating these rare events, straightforward MD simulations typically fail to sample the configuration space adequately \cite{genheden2012will}. Let us focus on the stochastic transitions between two metastable states $A \rightleftharpoons B$, which encapsulates the essential aspects of numerous molecular events and serves as a foundation for intricate multistep transitions \cite{martini2017variational}. A long ergodic trajectory, $\textbf{x}(t)$, would sample the equilibrium dynamics in the states, the rare excursions away from the states, and the even rarer transitions between states (Fig.~\ref{fig:1}A). After sampling many $A \rightleftharpoons B$ transitions, such trajectory would provide direct access to the Boltzmann distribution, $\rho(x)\propto \exp(-\beta U(x))$, where $U(x)$ is the potential energy of configuration $x$, and $\beta=1/k_\mathrm{B} T$; and to the mechanism, thermodynamics, and kinetics. In practice, obtaining many transitions for nontrivial systems is computationally unfeasible, as the simulation would hardly leave the metastable states and, therefore, would not explore the interesting dynamical bottleneck regions of the configuration space.
Moreover, relying on simulating a single (extremely) long trajectory for estimating observables 
is inefficient \cite{zuckerman2011equilibrium,ormeno2024convergence}.

The configuration space is sampled considerably faster by generating short unbiased trajectory segments (or paths) in convenient regions, terminating them, and then re-initiating them somewhere else while retaining correct kinetic and mechanistic information.  This realization gave rise to various path sampling algorithms \cite{prinz2011probing,bolhuis2002transition,chong2017pathsampling,zuckerman2017weighted}, which can be implemented, for instance, using Markov Chain Monte Carlo (MCMC) \cite{dellago1998transition,van2005elaborating}. This approach, however, presents challenges. How should we initiate those paths? How do we ensure they obey detailed balance? And how should we then combine the results together to obtain the same statistics we would get from a long ergodic trajectory?

We want to follow a ``divide and conquer'' strategy, initiating these paths in different regions of the configuration space, such that we separately sample the dynamics in the states, as well as representative excursions and transitions. To generate a path, we employ two-way shooting moves from transition path sampling  (TPS) \cite{bolhuis2002transition,bolhuis2021transition}: starting from shooting points (SPs)---initial configurations in the reactive region---we simulate backward and forward in time until reaching a state boundary, time-reverse the backward trajectory segment, and join the two parts together. The result of two-way shooting is a path passing through the SP following unbiased dynamics.

Suppose we could draw SPs directly from the Boltzmann distribution restricted to the reactive region. Realizing that the unbiased dynamics along the trajectory segments preserve the Boltzmann distribution, we could select a new SP from the previously generated path and automatically satisfy detailed balance. While we could then use these paths directly for our estimates, we would still disproportionately initiate trajectories near the boundaries of states A and B, with very infrequent sampling at the transition state (TS), and hence very few, if any, transition paths. 

Here, we build on the idea that we can enhance the sampling of the unlikely transition regions by employing a SP selection bias \cite{brotzakis2016one,jung2017transition}. 
The most ideal biasing function would allow for \textit{uniform} sampling between A and B.
Therefore, we propose to select SPs with a bias factor   
\begin{equation}\label{eq:bias}
    b(x) \sim  \exp(\beta F(\lambda(x))),  
\end{equation}
 where $F(\lambda)$ is the free energy profile along a reaction coordinate (RC) $\lambda (x)$.
This selection bias would balance the sampling in the reactive region in an optimal manner \cite{dinner2018trajectory}.
Note that
this is very different from 
other path sampling algorithms, as 
we do not  
{\it  enforce  a certain path distribution}, but  switched our perspective  to {\it sampling a  ensemble of SPs}, reducing the sampling problem  to constructing a Markov Chain of SPs following an adequate distribution.
This novel strategy requires: (\textbf{i}) access to  a good RC and (\textbf{ii}) its free energy profile,
(\textbf{iii}) a way to propose SPs, (\textbf{iv})  enforcing detailed balance in the Markov chain of these points, and (\textbf{v}) 
a method to properly reweight the trajectories to recover their equilibrium probabilities.

In this Letter, we propose a rejection-free path sampling (RFPS) algorithm ensuring  optimal sampling by addressing the five challenges listed above. We show how to build a Markov chain of SPs that satisfies detailed balance and, in this way, allows to  accept all paths generated from the SPs, including excursions and transition paths, without wasting computational resources. We additionally introduce a computationally efficient path reweighting strategy that allows us to retrieve the correct unbiased probabilities.  
We implement this algorithm by coupling it with an adaptive path sampling framework, which employs machine learning algorithms to learn an optimal reaction coordinate and use it to guide the sampling of rare paths \cite{jung2023machine,lazzeri2023molecular}. We demonstrate the algorithm's efficacy through analytical benchmarks and a molecular system application.


\noindent{\textit{Building a Markov Chain of SPs}} ---  A path $\textbf{x}(t) = \{\textbf{x}(0),~\textbf{x}(\Delta t),~\dots,~\textbf{x}(L[\textbf{x}]\Delta t)\}$ is a collection of frames connected by unbiased dynamics, where $\Delta t$ is the (fixed) time interval between frames and $L[\textbf{x}]$ is the (variable) path length. Paths are defined unambiguously based on the states: they all begin and end upon crossing a state boundary. With two metastable states $\mathrm{A}$ and $\mathrm{B}$, paths are of four types: (\textbf{i}) \textit{in-$\mathrm{A}$ segments}, with initial and final frame outside $\mathrm{A}$, and all other frames in $\mathrm{A}$; (\textbf{ii}) \textit{from-$\mathrm{A}$ excursions}, with initial frame in $\mathrm{A}$, final frame in either $\mathrm{A}$ or $\mathrm{B}$, and all other frames in the reactive region $\mathrm{R}=\overline{\mathrm{A}\cup\mathrm{B}}$; (\textbf{iii}) \textit{in-$\mathrm{B}$ segments}; and (\textbf{iv}) \textit{from-$\mathrm{B}$ excursions}. The excursions connecting two different states are called \textit{transitions} and form the transition path ensemble (TPE). The equilibrium path ensemble (PE) consists of all possible paths with their occurrence probabilities (Fig.~\ref{fig:1}A).

Large excursions and transitions are rare, as they need to overcome high energy barriers. To obtain these, our enhanced sampling strategy enforces a SP distribution
\begin{equation}\label{eq:P_sp}
P_{\mathrm{sp}}(x) =
\rho(x)~b(x),
\end{equation}
where $\rho(x)$ is the Boltzmann distribution and $b(x)$ is the selection bias. An adequate choice of $b(x)$ allows   uniform sampling of the SPs in $\mathrm{R}$ and thus uniform coverage of the associated paths---a realization of the ``divide and conquer'' approach described in the introduction.
To this purpose, we introduce a RC $\lambda(x)$ for the $\mathrm{A}$-to-$\mathrm{B}$ transition, and 
apply the bias given in  Eq.~\eqref{eq:bias}
between
two limiting isosurfaces $\lambda_{\mathrm{A}}$ and $\lambda_{\mathrm{B}}$ (close but \textit{not} identical to the state boundaries) and 0 otherwise.
We only bias the selection along $\lambda$: when restricting ourselves on a $\lambda$ isosurface (or \textit{interface}), the SPs remain Boltzmann-distributed. Clearly, Eq.~\eqref{eq:bias} requires knowledge of  $F(\lambda)$; we will show how to obtain it self-consistently by reweighting the sampled paths later in this Letter.

We now want to construct a Markov chain of SPs, using the following trial move $x \rightarrow y $: (\textbf{i}) generate a path $\textbf{x}$ by two-way shooting, (\textbf{ii}) assign selection probabilities to the frames of $\textbf{x}$ according to
\begin{equation}\label{eq:psel}
    p_{\mathrm{sel}}(y;\textbf{x}) = \frac{b(y)}{\sum_{j=1}^{L[\textbf{x}]-1} b(\textbf{x}(j\Delta t))},
\end{equation}
and (\textbf{iii}) select $y \in \textbf{x}$ according to $p_{\mathrm{sel}}$. Remarkably, by always accepting the move, the chain satisfies detailed balance and converges to $P_{\mathrm{sp}}(x )$ of Eq.~\eqref{eq:P_sp} (proof in the Supplemental Material~\cite{prl2025supplemental}).
Essentially, \textit{the distribution bias is also the selection bias}: we go from a standard MCMC algorithm with acceptance/rejection of trial elements to an algorithm where the target distribution is encoded by $b(x)$. Thus, we achieve uniform SP sampling across $\lambda$ reflected in frequent roundtrips in the reactive region \cite{bolhuis2008rare}. In the context of SPs connected by unbiased dynamics, uniform sampling is an optimal stratification strategy \cite{dinner2018trajectory}. Note that enhanced path sampling is a consequence of  the SP sampling. Similar to other path sampling algorithms such as TPS \cite{bolhuis2002transition}, we rely on the unbiased dynamics to decorrelate not only the SPs but also the associated paths. 

\noindent{\textit{Path reweighting}} --- With RFPS, we generate a chain of SPs in the reactive space with associated paths, where rare excursions and transitions are over-expressed compared to the equilibrium statistics (Fig.~\ref{fig:1}B). Furthermore, the PE also contains internal segments and very small excursions that are not typically obtained with MCMC path sampling. To cover for them, we complement our data with ``free'' unbiased simulations around the metastable states and cut the resulting trajectories at the state borders (Fig.~\ref{fig:1}C). To restore the correct equilibrium frequency, a reweighting algorithm downplays the large excursions and transitions \cite{rogal2010reweighted}.  The outcome is the reweighted path ensemble: 
\begin{equation}
\mathrm{RPE} = \left\{(w_1,~\textbf{x}_1),~\dots,~(w_n,~\textbf{x}_n)\right\},
\end{equation}
where every generated path $\textbf{x}_i$ has a weight $w_i$. The RPE is an estimate of the PE and provides direct access to free energy projections, transition mechanism, and rates estimates (see the Supplemental Material for details~\cite{prl2025supplemental}).

For reweighting the paths in an efficient way, we introduce a
continuous interface representation \cite{lazzeri2023molecular},
based on the RC value $\lambda$.
Specifically, for each SP $x_i$, we define the \textit{shooting interface}
\begin{equation}
    \lambda_i = \begin{cases}
        \lambda(x_i)&\Leftrightarrow~\textbf{x}_i~\text{from path sampling},\\
        -\infty&\Leftrightarrow~\textbf{x}_i~\text{from free simulations around}~\mathrm{A},\\
        +\infty&\Leftrightarrow~\textbf{x}_i~\text{from free simulations around}~\mathrm{B},
    \end{cases}
\end{equation}
and associate the corresponding path $\textbf{x}_i$ with $\lambda_i$ \cite{van2003novel,rogal2010reweighted}.
The selection bias of Eq.~\eqref{eq:bias} guarantees that, at convergence, the SPs are Boltzmann-distributed on their respective shooting interfaces. However, $\mathbf{x}_i$ can have more frames than just $x_i$ at the shooting interface, and in that case, the path is overrepresented \cite{falkner2023conditioning, hummer2004transition, bolhuis2008rare, falkner2024revisiting, frenkel2006wasterecycling,brotzakis2019approximating}. To restore the equilibrium importance of $\mathbf{x}_i$ at the shooting interface, we use
\begin{equation}
    f_i = 1/n_{\lambda_i}[\textbf{x}_i]
    \label{eq:fi}
\end{equation}
as a correction factor, where $n_{\lambda_i}[\textbf{x}_i]$, the (normalized) density of path frames at the shooting interface.
If $\textbf{x}_i$ comes from a free simulation, $n_{\pm\infty}[\textbf{x}_i]= 1$, and $f_i\equiv 1$. At convergence, the $f$-corrected paths uniformly sample the many
interfaces across $\lambda$ (see the Supplemental Material for details~\cite{prl2025supplemental}).

We reweight the excursions from either state and the internal segments within either state separately. Reweighting the excursions from one state involves matching the equilibrium {\it crossing probability} and the sampled crossing distribution \cite{rogal2010reweighted}. The crossing probability from A, $P_{\mathrm{A}}(\lambda)$, is the fraction of equilibrium excursions crossing interface $\lambda$; it is an ever-decreasing function with a minimum at the boundary of state B, $P_{\mathrm{A}}(\mathrm{B})$, corresponding to the frequency of occurrence of spontaneous transitions \cite{van2003novel,van2005elaborating,cabriolu2017foundations}. Similarly, $P_{\mathrm{B}}(\lambda)$ is the crossing probability 
for the equilibrium excursions from $\mathrm{B}$. The actual sampled distributions from our path sampling simulations are recorded in the histograms $m_{\mathrm{A}}(\lambda)$ and $m_{\mathrm{B}}(\lambda)$. 
Defining the extreme value $\Lambda_i$ of an excursion $\textbf{x}_i$ as the maximum $\lambda$  along paths from A, or the minimum  $\lambda$  along paths from B,
these unweighted histograms are  
\begin{subequations}
 \begin{align}
    m_{\mathrm{A}}(\lambda) &= \sum_{i=1}^n h_{\mathrm{R,A}}[\textbf{x}_i]~ f_i~\textbf{1}_{(\lambda_i,~\Lambda_i]}(\lambda),\\
    m_{\mathrm{B}}(\lambda) &= \sum_{i=1}^n h_{\mathrm{R,B}}[\textbf{x}_i]~ f_i~\textbf{1}_{[\Lambda_i,~\lambda_i)}(\lambda),
\end{align}   
\end{subequations}
where $h_{\mathrm{R,A}}$ and $h_{\mathrm{R,B}}$ are unity for excursions from $\mathrm{A}$ and $\mathrm{B}$, respectively, and zero otherwise.  $\textbf{1}_{(a,b)}(z)$ is an indicator function that is 1 if $z$ is between $a$ and $b$, and zero otherwise. 
We then reweight each excursion by dividing by the sampled crossing distribution and multiplying by the target crossing probability:
\begin{equation}\label{eq:excursions_weights}
\begin{aligned}
    w_i = f_i\Bigg[h_{\mathrm{R,A}}[\textbf{x}_i]~\frac{P_{\mathrm{A}}(\Lambda_i)}{m_{\mathrm{A}}(\Lambda_i)}+ h_{\mathrm{R,B}}[\textbf{x}_i]~\frac{n_{\mathrm{BA}}}{n_{\mathrm{AB}}}\frac{P_{\mathrm{B}}(\Lambda_i)}{m_{\mathrm{B}}(\Lambda_i)}\Bigg].
\end{aligned}
\end{equation}
%
The correct equilibrium kinetics is recovered by imposing an equal number of $\mathrm{A}$-to-$\mathrm{B}$ and $\mathrm{B}$-to-$\mathrm{A}$ transitions:
\begin{equation}
    n_{\mathrm{AB}}~P_{\mathrm{A}}(\mathrm{B}) = n_{\mathrm{BA}}~P_{\mathrm{B}}(\mathrm{A}).
\end{equation}
Finally, the segments internal to the metastable states have weight
\begin{equation}\label{eq:internal_weights}
    w_i = h_{\mathrm{A}}[\textbf{x}_i]~c_{\mathrm{A}} +  h_{\mathrm{B}}[\textbf{x}_i]~c_{\mathrm{B}},
\end{equation}
where $h_{\mathrm{A}}$ and $h_{\mathrm{A}}$ select the in-$\mathrm{A}$ and in-$\mathrm{B}$ segments, respectively, and the $c_{\mathrm{A}}$ and $c_{\mathrm{B}}$ normalization constants match the fluxes in and out of the states:
\begin{align}
    \sum_i c_{\mathrm{A}}~h_{\mathrm{A}}[\textbf{x}_i] =  \sum_i h_{\mathrm{R,A}}[\textbf{x}_i]~w_i.
    \label{eq:cAmatch}
    \end{align} 
While an iterative solution for the  crossing probability is possible using Eqs.~\ref{eq:fi}-\ref{eq:cAmatch}, in the Supplemental Material we provide a maximum-likelihood closed solution for reweighting the paths  and computing the equilibrium crossing probabilities 
\cite{prl2025supplemental}.

\noindent{\textit{Practical implementation: RFPS-AIMMD}} --- Implementing our {\color{black}enhanced sampling} algorithm involves two main challenges: obtaining a good RC and accurately estimating the free energy along the RC. Both are essential for proper SP selection through Eq.~\eqref{eq:bias}. Why is the RC so important? While the proposed sampling and reweighting algorithms are optimal for a given RC, a good RC further accelerates convergence. We argue that choosing
\begin{equation}
    \lambda(x) = \log\frac{p_{\mathrm{B}}(x)}{1-p_{\mathrm{B}}(x)} \equiv \mathrm{logit} ~p_{\mathrm{B}}(x),
\end{equation}
$\forall \lambda(x)\in[\lambda_{\mathrm{A}},\lambda_{\mathrm{B}}]$, where $p_{\mathrm{B}}(x)$ is the committor of the $\mathrm{A}$-to-$\mathrm{B}$ transition (the optimal RC \cite{banushkina2016optimal}), guarantees both maximum diffusion in $\mathrm{R}$ and optimal reweighting performance. From here on, we will refer to $\lambda$ as ``logit committor.''

AI for Molecular Mechanism Discovery (AIMMD) can address both the RC learning and $F(\lambda)$ estimation challenges simultaneously \cite{jung2023machine, lazzeri2023molecular,post2025aiguidedtransitionpathsampling,Horvath2025.03.09.638703}. Thus, it is a natural choice for optimizing the bias and thus the sampling on the fly in a data-driven way. In AIMMD, a neural network (NN) learns a model of the committor $\tilde p_{\mathrm{B}}(x)$, approximating the true committor $p_{\mathrm{B}}(x)$, and uses it to guide SP selection \cite{jung2023machine}. The learning process involves maximizing the likelihood of the ``shooting results'' $(r_{\mathrm{A},i}, r_{\mathrm{B},i})$, i.e., the number of times shooting from $x_i$ reached states A and B, respectively \cite{peters2006obtaining}. This gives the training loss:
\begin{equation}\label{eq:loss}
    \mathcal L = -\sum_{i=1}^n r_{\mathrm{A},i}~\log (1-\tilde p_{\mathrm{B}}(x_i)) + r_{\mathrm{B},i}~\log \tilde p_{\mathrm{B}}(x_i).
\end{equation}

AIMMD improves both exploration and accuracy \cite{lazzeri2023molecular} since the NN can capture the significant nonlinearity of the committor in Cartesian space \cite{jung2023machine, hartmann2013characterization}. Additionally, it allows for path reweighing and free energy estimates in real time. While the original AIMMD algorithm used TPS with waste-recycling of the trial paths \cite{brotzakis2019approximating, lazzeri2023molecular}, here we integrate it with RFPS.
In RFPS-AIMMD, the bias changes over time but iteratively converges to the optimal solution. We summarize the full algorithm in the following steps (see also Fig.\ref{fig:1}D and the Supplemental Material \cite{prl2025supplemental}).
\begin{itemize}
    \item[{0}] \textit{{Initialization}}: start with an initial transition $\textbf{x}_0$ and committor trained on a few frames in $\mathrm{A}$ and $\mathrm{B}$.
    \item[{1a}] \textit{{Path sampling}}: generate path $\textbf{x}_i$ as part of the SP move $x_{i} \rightarrow x_{{i+1}}$ in the Markov chain.
    \item[{1b}] \textit{{Free sampling}}: run unbiased simulations around the states; generate internal segments and short excursions by cutting the resulting trajectories.
    \item[{2}] \textit{{Committor learning}}: once obtained $\textbf{x}_i$, relearn the committor $\lambda(x)$ by minimizing the loss in Eq.~\ref{eq:loss} (see also Refs.~\cite{lazzeri2023molecular,jung2023machine}).
    \item[{3}] \textit{{PE estimate}}: collect all  generated paths and reweight them based on $\lambda$; estimate $F(\lambda)$.
    \item[{4}] \textit{{SP selection}}: optimize the $\lambda_{\mathrm{A}}$ and $\lambda_{\mathrm{B}}$ boundaries based on the sampling performance; update $b(x)$; select SP $x_{i+1}$ from $\textbf{x}_i$ and update the chain.
    \item[]\textit{Repeat steps 1-4 until the estimates converge}.
\end{itemize}


\begin{figure}
\includegraphics[width=0.45\textwidth]{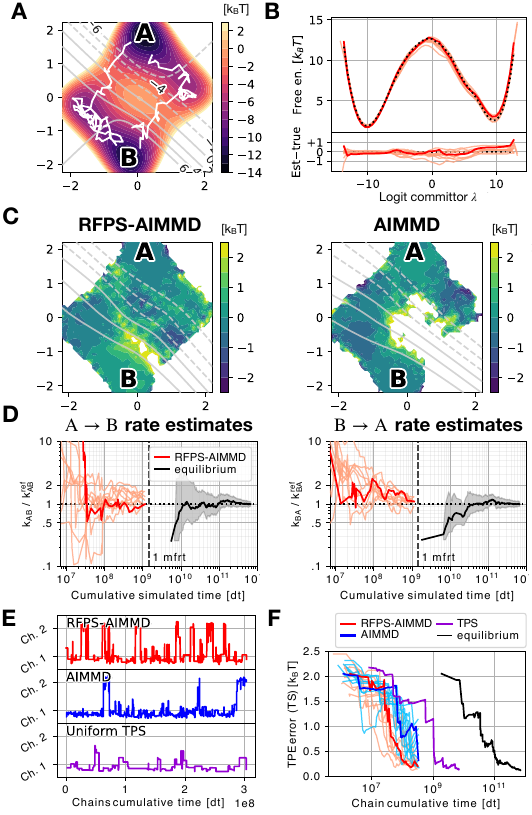}
\caption{\label{fig:2} Wolfe-Quapp system.
\textbf{(A)} Potential energy surface with highlighted the states $\mathrm{A}$ and $\mathrm{B}$. Two representative transitions in the lower (left) and higher-energy (right) reactive channels are superimposed on the contour plot (white lines). The logit committor $\lambda$ (gray isolines) comes from the numerical solution of the stationary Fokker-Plank equation.
\textbf{(B)} RFPS-AIMMD with a fixed budget of $1.5 \times 10^8~[dt]= 0.1~$mfrt, free energy estimates of ten independent runs (first run: red; other runs: light red) projected on the $\lambda$ learned in the first run. The dotted line is the true free energy.
\textbf{(C)} Enhanced sampling runs with a fixed budget of $0.1~$mfrt, error in the 2D free energy with two different sampling strategies: RFPS-AIMMD (left) and AIMMD (right). We show the results of the first out of 10 independent runs for each strategy. The learned $\lambda$ is superimposed on the contour plot for each run (gray isolines).
\textbf{(D)} Evolution of the transition rates estimates during the ten RFPS-AIMMD runs (first run: red; other runs: light red). The mfrt scale and the results from a long equilibrium run (black) are shown for reference. The gray area is the $95\%$ confidence interval of the equilibrium estimate.
\textbf{(E)} Time series of the reactive channels distance of the latest sampled transition in the first RFPS-AIMMD (top) and AIMMD runs (center) and in a standard TPS run (bottom).
\textbf{(F)} Evolution of the TS root-mean-square error (RMSE) from the TPE sampled in RFPS-AIMMD (first run: red; other runs: light red), AIMMD (first run: blue; other runs: light blue), TPS (violet), and long equilibrium runs (black).
}
\end{figure}

\begin{figure}
\includegraphics[width=0.45\textwidth]{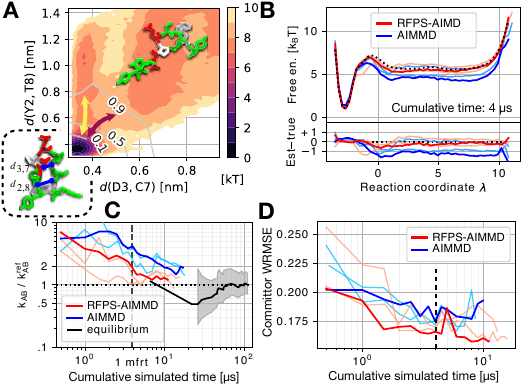}
\caption{\label{fig:3} Chignolin.
\textbf{(A)} Reference free energy from the long equilibrium runs projected onto the $(d_{3,7}, d_{2,8})$ plane. Representative renders of the folded (bottom left) and unfolded conformation (top right) are superimposed on the plot, highlighting the $d_{3,7}$ and $d_{2,8}$ CVs. The effective committor (gray isolines) is computed from the PE, as in Ref.~\cite{bolhuis2011relation}. The ``zipping'' and ``hydrophobic collapse'' folding/unfolding pathways are indicated by yellow and dark red arrows, respectively.
\textbf{(B)} Enhanced sampling with a fixed budget of $4~\text{µs}\approx 1\mathrm{mfrt}$, free energy estimates from the three RFPS-AIMMD runs (first run: red; other runs: light red) projected onto the $\lambda$ learned during the first run. The estimates from the three AIMMD runs are also included (first run: blue; other runs: light blue). The dotted line is the reference free energy from the long equilibrium runs.
\textbf{(C)} Evolution of the unfolding rates estimates during the three RFPS-AIMMD (first run: blue; other runs: light blue) and the AIMMD runs (first run: red; other runs: light red) given the cumulative simulated time. The mfrt scale and results from a long equilibrium run (black) are included for reference. The gray area is the $95\%$ confidence interval of the equilibrium estimate.
\textbf{(D)} Evolution of the committor WRMSE on the validation set for both the RFPS-AIMMD runs (first run: red; other runs: light red) and the AIMMD runs (first run: blue; other runs: light blue).
}
\end{figure}

\noindent{\textit{Results}} --- We  tested RFPS-AIMMD on two bi-stable, asymmetric systems:
\textbf{(i)} the Wolfe-Quapp 2D potential (WQ) has two reactive channels with $\approx 11~k_{\mathrm{B}}T$ and $12~k_{\mathrm{B}}T$ energy barriers (Fig.~\ref{fig:2}A) \cite{quapp2004reaction};
\textbf{(ii)} chignolin (PDB: CLN025) in water shows nontrivial folding/unfolding dynamics with multiple folding pathways, a significant energy gap, and is a fitting introduction to more complex biomolecular processes \cite{satoh2006folding,lindorff2011fastfolding,kato2015nmr,zschau2023mechanism} (Fig.~\ref{fig:3}A). In each run, we allocated one computing node for path sampling, one for free simulations around $\mathrm{A}$, and one for free simulations around $\mathrm{B}$. Additionally, we ran long equilibrium simulations yielding reference transition rates and, in the case of chignolin, the Boltzmann distribution. 
Computational details  can be found in the Supplemental Material \cite{prl2025supplemental}.

For the WQ system, we collected ten independent RFPS-AIMMD and ten AIMMD runs, starting from the low-energy channel. To evaluate our method in a data-scarce regime, we first analyzed the results after a computational budget of \(\tau_{\max} = 1.5 \times 10^8\) integration steps, or $0.1$ mean first return time (mfrt). The mfrt is the average time it takes for an equilibrium simulation to produce two consecutive transitions, providing the scale of exploration without enhanced sampling \cite{berezhkovskii2019committors,benichou2015mean}.

RFPS-AIMMD enabled a very fast sampling that quickly discovered the presence of both channels, and populated them with the right proportions. This resulted in a more accurate  
 free energy estimates compared to the already very efficient AIMMD (Figs.~\ref{fig:2}C,~S6).
As a consistency check, the free energy profiles calculated along the committor were within $0.5~k_{\mathrm{B}}T$ of the reference (Figs.~\ref{fig:2}B,~S6C) for nearly all runs.
Increasing $\tau_{\max}$ resulted in improved  energy estimates and flattened SP selection histograms along $\lambda$ (Fig.~S7).

A quick equilibration in $\mathrm{R}$ manifests in the frequency of reactive channel switches: the faster transitions alternate between channels, the faster they decorrelate. Each RFPS-AIMMD run significantly improved the switch frequency over the AIMMD runs and standard TPS (Figs.~\ref{fig:2}E,~S8, and Table~I in the Supplemental Material \cite{prl2025supplemental}). The higher frequency accelerated the convergence of the TS from the sampled TPE (Fig.~\ref{fig:2}F), disclosing the transition mechanism. Additionally, it sped up committor learning in both channels (see Fig.~\ref{fig:2}C). The smaller excursions in the shooting chains play a crucial role in increasing the switch frequency, as the reactive channels are least separated close to the metastable states. By consistently selecting new SPs also from these paths, we significantly enhance exploration without compromising exploitation. As a result, we perform better than standard TPS, even if the focus is on obtaining all paths and not just transitions.

We tracked the evolution of the transition rates estimates as a function of the increasing amount of simulated time (Fig.~\ref{fig:2}D). All runs converged within a factor $2$ of the reference rates at around $0.1$ mfrt, orders of magnitude before the equilibrium results. However, poorer learning performances near the states tend to overestimate the crossing probability and thus the rates, introducing a systematic error \cite{ghysbrecht2023accuracy}.


In the case of chignolin, the three RFPS-AIMMD runs outperformed the AIMMD runs in both free energy and rate estimates (see Fig.~\ref{fig:3}B,C, Figs.~S10,~S11, and Table II in the Supplemental Material). Only the new method consistently achieved unfolding rate estimates (which are the rarest) within a factor $2$ of the reference in less than $4~\text{µs} \approx 1~\mathrm{mfrt}$ computational budget. Compared to WQ, the lower computational gain compared to equilibrium simulations is due to the low energy barrier, less precise state definitions, and longer paths in terms of mfrt \cite{lazzeri2023molecular}.

When assessed on an external validation set, the committor models from RFPS-AIMMD were more accurate than those from the original AIMMD algorithm, particularly near state $\mathrm{A}$ (Figs.~S15D). Notably, this discrepancy revealed a positive correlation between the errors in the committor estimates and the unfolding rate throughout the simulations (Fig.~\ref{fig:3}D). We believe that RFPS provides a better SP coverage of $\mathrm{R}$ and thus enhances both committor learning and path sampling in a complementary manner. Thus, the learned models generalize more effectively and give better insight. This is particularly important when multiple main reactive channels are present, such as the ``zipping'' and ``hydrophobic collapse'' pathways \cite{zschau2023mechanism} illustrated in Fig.~\ref{fig:3}A. However, the lack of suitable CVs that are orthogonal to the committor limited our ability to assess the decorrelation speed of the two methods.

\noindent{\textit{Discussion}} --- We introduced RFPS, an enhanced path sampling scheme where the Markov chain elements are SPs, and the paths are the products of the SP moves. No simulations are wasted: by appropriately reweighting the paths, we can recover the equilibrium PE along with accurate free energy projections, transition mechanisms, and rates. We combined RFPS with AIMMD to iteratively learn the committor and use it to optimize sampling.

The idea behind our sampling scheme is that we can perform efficient and optimal path sampling simulation just by controlling the SP distribution, as previously proposed by Falkner et al. \cite{falkner2023conditioning}. To this purpose, the SPs do not necessarily have to come from a MCMC. If a Boltzmann generator \cite{noe2019boltzmann}, or equivalently an already converged PE estimate, is available, one could use it for sampling many SPs at the same time and run embarrassingly parallel simulations. However, RFPS-AIMMD is optimal for quick exploration of the configuration space when little prior information is available and provides a theoretically rigorous solution to the sampling problem. We  stress that one does not have to necessarily bias along the RC, even though this solution allows for the robust reweighting algorithm that we propose in this Letter. As long as $b(x)$ is defined in the reactive region, the SPs from RFPS will eventually converge to the distribution of Eq.~\eqref{eq:P_sp}. This characteristic allows for tailored solutions, e.g., limiting the sampling in one reactive channel, or promoting channel switches by adding additional biasing terms orthogonal to the committor as in Wang-Landau (WL) sampling \cite{du2013adaptive}.

The RFPS algorithm by itself
bears interesting similarities to well-tempered metadynamics \cite{barducci2008well,dama2014well}. A similar conceptual switch from biasing potentials to probability distributions is what led to OPES \cite{invernizzi2020rethinking}. In both approaches, the bias is updated iteratively based on previous simulations, which yield equilibrium estimates. This process ensures internal consistency and establishes a convergence criterion: when the bias converges, the sampling converges, thus validating the free energy projection on $\lambda$. However, we emphasize that in our case the PE gives direct access to complete thermodynamic and kinetic information. RFPS also has strong connections with adaptive Single Replica TIS \cite{du2013adaptive,cabriolu2017foundations}, where the WL bias function plays the same role as 
our selection bias \cite{du2013adaptive}.

RFPS-AIMMD has proven to be highly effective in accelerating the exploration of reactive spaces and providing quantitative estimates compared to the original AIMMD algorithm, TPS, and equilibrium simulations. With chignolin, the better training set coverage consistently improved the committor model in synergy with other results. We expect coverage to be even more critical as we raise the complexity of the systems.

RFPS-AIMMD demonstrated notable robustness and convergence speed. This is ultimately because paths follow the unbiased dynamics: any inaccuracies made in the SP selection would be partially corrected through the shooting simulations. As a result, we were able to keep the early data generated from non-converged models and still improve our estimates. This self-healing property is exclusive to using unbiased dynamics. One can also inject a better committor model \textit{a posteriori} and further improve the results without additional sampling.

The RFPS-AIMMD algorithm is parallelizable, as multiple RFPS chains and free simulations can run at the same time \cite{jackel2024free,jung2023machine}. Like TPS and TIS, it performs better for systems with high energy barriers and short-lived intermediates. However, barriers that are too high barriers can lead to situations where the SPs become trapped near metastable states, hindering convergence. A careful definition of the state boundaries can mitigate this problem. One solution is to redefine the chain to allow multiple paths to be available simultaneously for SP selection. Another option is to occasionally override the SP selection by allowing configurations from free simulations, which can further accelerate exploration. Additionally, one can augment the committor's training set by including all the configurations in the PE. 

In conclusion, we present a novel adaptive and data-efficient path sampling algorithm that enables the characterization of challenging rare molecular events.

\textit{Acknowledgements}~---~We thank Attila Szabo for the stimulating discussions. 
G.L. and R.C. acknowledge the support of Goethe
University Frankfurt, the Frankfurt Institute for
Advanced Studies, the LOEWE Center for Multiscale
Modelling in Life Sciences of the state of Hesse, the
CRC 1507: Membrane-associated Protein Assemblies,
Machineries, and Supercomplexes (P09), and the In-
ternational Max Planck Research School on Cellular
Biophysics. P.G.B. acknowledges the support of the University of Amsterdam.
We than the Center for Scienfitic Computing of Goethe-University for computational support. All data were generated on the Goethe HLR cluster.

\textit{Data availability}~---~The data and code from this study are freely available in the Zenodo repository of Ref.~\cite{prl2025zenodo}. 

\bibliographystyle{apsrev4-2}
\bibliography{arxiv_250326}
\end{document}


\title{Supplemental Material for ``Optimal Rejection-Free  Path Sampling ''}

\author{Gianmarco Lazzeri}
\email{lazzeri@fias.uni-frankfurt.de}
\affiliation{Institute of Biochemistry, Goethe University Frankfurt}
\affiliation{Frankfurt Institute for Advanced Studies}

\author{Peter G. Bolhuis}
\email{P.G.Bolhuis@uva.nl}
\affiliation{Van 't Hoff Institute for Molecular Sciences, University of Amsterdam}

\author{Roberto Covino }
\email{covino@fias.uni-frankfurt.de}
\affiliation{Institute of Computer Science, Goethe University Frankfurt}
\affiliation{Frankfurt Institute for Advanced Studies}

\date{\today}



\maketitle

\section{The (Reweighted) Path Ensemble}

In this Section, we show how we can estimate free energies and transition rates from the equilibrium path ensemble (PE). We define our thermodynamic and kinetic estimates as averages over an infinitely long, ergodic MD trajectory. We start from the PE extracted from equilibrium simulations (where all paths have the same weight) and generalize to the case of unequal weights (the reweighted path ensemble, RPE).

Assume we can partition the configuration space in $\mathrm{A}$, $\mathrm{B}$, \dots~metastable states and the reactive region $\mathrm{R}$ in between (Fig.~\ref{fig:S1}A). As explained in the Main Text, we can split an equilibrium trajectory in $\textbf{x}_1,~\dots,~\textbf{x}_n$ paths, which sample the PE. Those paths would be either internal segments or excursions, and a tiny fraction of the excursions would be transitions. For each path \textbf{x}, the first and last frames $\textbf{x}(0),~\textbf{x}(L[\textbf{x}]\Delta t)$ are \textit{boundary frames}. From our definition of paths, the boundary frames would belong to different regions of the configuration space than the \textit{internal frames} $\textbf{x}(\Delta t),~\textbf{x}(2\Delta t),~\dots,~\textbf{x}((L[\textbf{x}]-1)\Delta t)$. For example, if $\textbf{x}$ is an in-A segment, the boundary frames would be in $\mathrm{R}$; if $\textbf{x}$ is a from-A excursion, the first frame would be in A, and the last would be either in A or in another metastable state (in the latter case, $\textbf{x}$ is also a transition). As a consequence, the boundary frames of subsequent paths extracted from the same trajectory would overlap with each other. For this reason, we must exclude the boundary frames when computing the thermodynamic and kinetic averages.

\subsection{Thermodynamic Averages and Free Energy Projections}

Given an observable $O(x)$ (a function of the configurations), we can compute its average over an infinitely long ergodic trajectory. Let $\textbf{x}_1,~\dots,~\textbf{x}_n$ be paths composing the PE. Then, each path would contribute to the average with its internal frames:
\begin{equation}\label{eq:O1}
    \langle O \rangle = \frac{\sum_{i=1}^n\sum_{j=1}^{L[\textbf{x}]-1} O(\textbf{x}_i(j\Delta t))}{\sum_{i=1}^n L[\textbf{x}]-1}.
\end{equation}
With this approach, we can compute equilibrium densities projected on arbitrary CV representations $q(x)$. In this special case, we must use the density operator $\delta(\cdot)$, a discretization of the Dirac delta:
\begin{equation}\label{eq:rho1}
    \rho(q') \propto  \sum_{i=1}^n\sum_{j=1}^{L[\textbf{x}]-1} \delta(q(\textbf{x}_i(j\Delta t))) - q').
\end{equation}
In practice, we define a grid $q$ space, and for each grid cell, we calculate the number of PE configurations inside the cell. The associated free energy comes from the Boltzmann inversion:
\begin{equation}\label{eq:F1}
    F(q) = -k_{\mathrm{B}}T~\log(\rho(q)).
\end{equation}

If the PE is derived from an (enhanced) path sampling campaign like RFPS-AIMMD instead of a long equilibrium trajectory, the above remarks are still valid. In updating Eqs.~\eqref{eq:O1},~\eqref{eq:rho1}, and \eqref{eq:F1}, we must assign the right weight to the sampled paths:
\begin{equation}\label{eq:O2}
    \langle O \rangle = \frac{\sum_{i=0}^n\sum_{j=1}^{L[\textbf{x}]-1} O(\textbf{x}_i(j\Delta t))~w_i}{\sum_{i=1}^n (L[\textbf{x}]-1)~w_i}.
\end{equation}
\begin{equation}\label{eq:rho2}
    \rho(q') \propto \sum_{i=1}^n\sum_{j=1}^{L[\textbf{x}]-1} \delta(q(\textbf{x}_i(j\Delta t))) - q')~w_i.
\end{equation}

\subsection{Transition Rate Constants}

\begin{figure}
\begin{minipage}{0.42\textwidth}
\textbf{(A)} Paths committed to 3 states
\includegraphics[width=\textwidth]{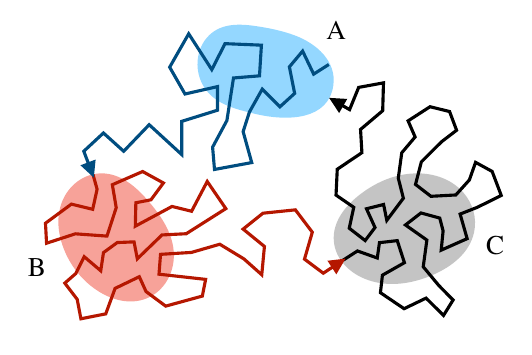}
\end{minipage}
\begin{minipage}{0.42\textwidth}
\textbf{(B)} Mfpt computation
\includegraphics[width=\textwidth]{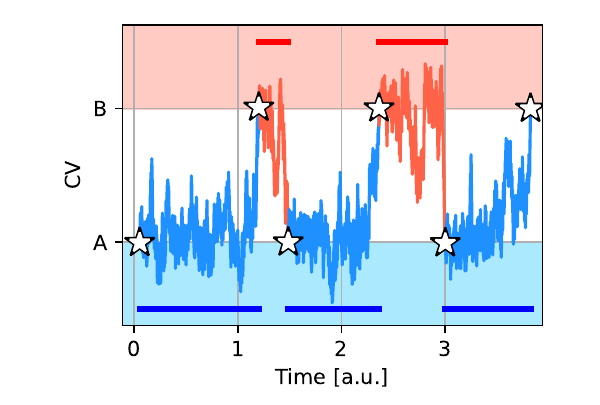}
\end{minipage}
    \caption{Transition rates estimate. \textbf{(A)} Schematics of an equilibrium trajectory in a configuration space with three metastable states ($\mathrm{A}$: blue, $\mathrm{B}$: red, and $\mathrm{C}$: gray), with the paths colored according to the committed state. \textbf{(B)} Time series of an unbiased trajectory in case of two metastable states, projected on the collective variable that defines the state boundaries. The transitions are completed at the white stars. The inverse of $k_{\mathrm{A}\mathrm{B}}$ (the mfpt or average waiting (residence) time $\tau_{\mathrm{A}\mathrm{B}}$) is the average duration of the blue segments, while the inverse of $k_{\mathrm{B}\mathrm{A}}$ is the average duration of the red segments. The mean first return time is the average waiting time for two consecutive transitions, i.e., the average interval between three consecutive transitions.}
    \label{fig:S1}
\end{figure}

We define the transition rate constants as ensemble averages over an infinitely long ergodic trajectory. In-$\mathrm{A}$ segments and from $\mathrm{A}$ excursions are both committed to $\mathrm{A}$. Let
\begin{equation}\label{eq:nAB}
    n_{\mathrm{A}\mathrm{B}} = \sum_{i=1}^n h_{\mathrm{A}\mathrm{B}}[\textbf{x}_i]
\end{equation}
be the total number of transitions from $\mathrm{A}$ to $\mathrm{B}$ in the PE, and
\begin{equation}\label{eq:TA}
    \mathcal{T}_{\mathrm{A}} = \sum_{i=1}^n h_{\mathrm{A}\mathrm{B}}[\textbf{x}_i]~(L[\textbf{x}_i]-1)~\Delta t
\end{equation}
the cumulative time of the paths committed to $\mathrm{A}$ (excluding the boundary frames). The transition rate constant,
\begin{equation}\label{eq:rate}
    k_{\mathrm{A}\mathrm{B}} = \frac{n_{\mathrm{A}\mathrm{B}}}{\mathcal{T}_{\mathrm{A}}},
\end{equation}
is the frequency of the $\mathrm{A}$-to-$\mathrm{B}$ transitions (in the following we will denote the rate constant simply as the rate). $\tau_{\mathrm{A}\mathrm{B}}=k_{\mathrm{A}\mathrm{B}}^{-1}$ is the mean first passage time (mfpt) from $\mathrm{A}$ to $\mathrm{B}$. In case of only two defined states, $\tau_{\mathrm{A}\mathrm{B}}$ becomes the average elapsed time between two subsequent transitions, the first arriving in $\mathrm{A}$ and the second in $\mathrm{B}$ (Fig.~\ref{fig:S1}B), a popular definition for the inverse rates \cite{peters2017reaction}. The mean first return time is $\mathrm{mfrt}=\tau_{\mathrm{A}\mathrm{B}}+\tau_{\mathrm{B}\mathrm{A}}$. The advantage of the proposed definition is that it can be used to build Markov models with an arbitrary number of states.

If the PE is derived from the AIMMD sampling campaign instead of a long equilibrium trajectory, the above remarks are still valid. In updating Eqs.~\eqref{eq:nAB} and \eqref{eq:TA}, we must assign the right weight to the sampled paths:
\begin{equation}\label{eq:nAB2}
    n_{\mathrm{A}\mathrm{B}} = \sum_{i=1}^n h_{\mathrm{A}\mathrm{B}}[\textbf{x}_i]~w_i,
\end{equation}
\begin{equation}\label{eq:TA2}
    \mathcal{T}_{\mathrm{A}} = \sum_{i=1}^n h_{\mathrm{A}\mathrm{B}}[\textbf{x}_i]~(L[\textbf{x}_i]-1)~w_i~\Delta t.
\end{equation}
By combining Eqs.~\eqref{eq:nAB2}, \eqref{eq:TA2}, and \eqref{eq:rate}, we obtain
\begin{subequations}
\begin{align}
    k_{\mathrm{AB}} = \Delta t^{-1}~\frac{\sum_{i=1}^n h_{\mathrm{A}\mathrm{B}}[\textbf{x}_i]~w_i}{\sum_{i=1}^n h_{\mathrm{A}\mathrm{B}}[\textbf{x}_i]~(L[\textbf{x}_i]-1)~w_i},\\
    k_{\mathrm{BA}} = \Delta t^{-1}~\frac{\sum_{i=1}^n h_{\mathrm{B}\mathrm{A}}[\textbf{x}_i]~w_i}{\sum_{i=1}^n h_{\mathrm{B}\mathrm{A}}[\textbf{x}_i]~(L[\textbf{x}_i]-1)~w_i}.
\end{align}
\end{subequations}
which we used to calculate the rates in this work, both for the reference equilibrium and RFPS-AIMMD estimates. This shows the consistency of this approach. We define the free energy difference between the two states as:
\begin{equation}
    \Delta F_{\mathrm{AB}}=F_{\mathrm{B}}-F_{\mathrm{A}}=-k_{\mathrm{B}}T~\log k_{\mathrm{AB}}/k_{\mathrm{BA}}.
\end{equation}

Note that this formulation is equivalent to the more traditional way of computing the rate constant in terms of  the flux through the first interface (which we assume to be the state boundary for simplicity) and crossing probability in the limit of average duration of a transition much smaller than the mfpt \cite{van2003novel}. The flux is simply $\phi= n_{\mathrm{A}}/\mathcal{T_\mathrm{A}}$, i.e., the number of paths $n_{\mathrm{A}}$ leaving A per unit time, whereas the crossing probability is $P_\mathrm{A}(\mathrm{B}) = n_{\mathrm{AB}}/n_\mathrm{A}$, so that the product is $k_{AB}= n_{AB}/\mathcal{T_\mathrm{A}}$ as expected.

The transition rates depend both on the state definitions and on the time resolution of the paths. Too indulgent state definitions lead to spurious transitions and increased (incorrect) rates. Too coarse time resolution leads to unregistered transitions and lower rates. Suboptimal collective variables fail at characterizing the metastable states, altering the perceived dynamics. In general, the correct results can only be meaningful if we succeed at problem specification.


\section{The SP Move Scheme Naturally Preserves Detailed Balance}

We now prove that the SP move $x\rightarrow y$ described in the Main Text preserves detailed balance.
The SP move implies doing two-way shooting from $x$ and generating the path $\textbf{x}$. This is only one of the many paths that could have been generated by two-way shooting from $x$. Those many paths follow the conditional distribution $\pi[\textbf{x};x]$. Let $s[\textbf{x}]$ be the frame index of $x$ within $\textbf{x}$ (the ``shooting index''):
\begin{equation}
    x = \textbf{x}(s[\textbf{x}]\Delta t).
\end{equation}
Then,
\begin{equation}
\begin{aligned}
\label{eq:conditional}
    \pi[\textbf{x};x] = h_{\mathrm{R}}[\textbf{x}]\prod_{j=s[\textbf{x}]}^{L[\textbf{x}]-1} p_{\Delta t}(\textbf{x}(j\Delta t)\rightarrow\textbf{x}((j+1)\Delta t))\\
   \times \prod_{j=0}^{s[\textbf{x}]-1}\overline{p}_{\Delta t}(\textbf{x}((j+1)\Delta t)\rightarrow\textbf{x}(j\Delta t))&,
\end{aligned}
\end{equation}
where $p_{\Delta t}(\mathbf{x}(j\Delta t)\rightarrow \mathbf{x}((j+1)\Delta t))$ is the $\Delta t$-propagator, $\overline p_{\Delta t}(\mathbf{x}((j+1)\Delta t)\rightarrow \mathbf{x}(j\Delta t))$ is the backward $\Delta t$-propagator, and $L[\mathbf{x}]$ is the path length. The indicator functional $h_{\mathrm{R}}[\textbf{x}]$ selects only excursions (with boundary frames inside the metastable states), i.e., is unity only for excursions, and zero otherwise. Eq.~\eqref{eq:conditional} combines the probabilities of the forward and backward simulations. Much like in two-way shooting, we can ``time-reverse'' the backward segment by repeatedly applying microscopic time reversibility:
\begin{equation}\label{eq:reversibility}
\begin{aligned}
    \rho(\textbf{x}(j\Delta t))~{p}_{\Delta t}(\textbf{x}(j\Delta t)\rightarrow\textbf{x}((j+1)\Delta t))&=\\\rho(\textbf{x}((j+1)\Delta t))~\overline{p}_{\Delta t}(\textbf{x}((j+1)\Delta t)\rightarrow\textbf{x}(j\Delta t))&,
\end{aligned}
\end{equation}
and obtain
\begin{equation}
\label{eq:conditional2}
    \pi[\textbf{x};x] = h_{\mathrm{R}}[\textbf{x}]~\frac{\rho(\textbf{x}(0))}{\rho(x)}\prod_{j=0}^{L[\textbf{x}]-1} p_{\Delta t}(\textbf{x}(j\Delta t)\rightarrow\textbf{x}((j+1)\Delta t)).
\end{equation}
Eq.~\eqref{eq:conditional2} reveals a simple relation between $\pi[\textbf{x};x]$ and the subset of the PE containing only the excursions, $\mathcal{P}_{\mathrm{R}}[\textbf{x}]$. Specifically:
\begin{equation}\label{eq:PR}
    \mathcal{P}_{\mathrm{R}}[\textbf{x}] = \mathcal Z^{-1}\times h_{\mathrm{R}}[\textbf{x}]~\rho(\textbf{x}(0))\prod_{j=0}^{L[\textbf{x}]-1} p_{\Delta t}(\textbf{x}(j\Delta t)\rightarrow\textbf{x}((j+1)\Delta t)),
\end{equation}
which injected in Eq.~\eqref{eq:conditional2} leads to
\begin{equation}\label{eq:pi_PE_relation}
\pi[\textbf{x};x] = \mathcal Z \times\frac{\mathcal{P}_{\mathrm{R}}[\textbf{x}]}{\rho(x)}~h_{x}[\textbf{x}],
\end{equation}
where $h_{x}[\textbf{x}]$ selects only the excursions that have $x$ as one of their frames. We emphasize that the normalization constant $\mathcal Z = \int \mathcal D\textbf{x}~\mathcal P_{\mathrm{R}}[\textbf{x}]$ is independent on $x$.

Generating $y$ involves selecting it from $\textbf{x}$. Thus, the generation probability
\begin{equation}
p_{\mathrm{gen}}(x\rightarrow y)=\int\mathcal{D}\textbf{x}~\pi[\textbf{x};x]~h_{x,y}[\textbf{x}]~p_{\mathrm{sel}}(y;\textbf{x})
\end{equation}
is a path integral over $\pi[\textbf{x};x]$, where only the paths that connect $x$ and $y$ (selected with $h_{x,y}$) contribute to $p_{\mathrm{gen}}$ proportionally to the selection probability $p_{\mathrm{sel}}(y;\textbf{x})$. Eq.\eqref{eq:pi_PE_relation} and the observation that $h_{x,y}h_{x}=h_{x,y}$ allows us to rewrite
\begin{equation}\label{eq:pgen}
\begin{aligned}
p_{\mathrm{gen}}(x\rightarrow y)&=\mathcal Z \times \int \mathcal{D}\textbf{x}~\frac{\mathcal{P}_{\mathrm{R}}[\textbf{x}]~h_{x,y}[\textbf{x}]~p_{\mathrm{sel}}(y;\textbf{x})}{\rho(x)}
\\&=\mathcal Z \times \frac{b(y)}{\rho(x)}\int \mathcal{D}\textbf{x}~\frac{\mathcal{P}_{\mathrm{R}}[\textbf{x}]~h_{x,y}[\textbf{x}]}{\sum_{j=1}^{L[\textbf{x}]} b(\textbf{x}(j\Delta t))},
\end{aligned}
\end{equation}
where we also made explicit $p_{\mathrm{sel}}$ of Eq.~(3) of the Main Text. The constant $b(y)/\rho(x)$ is now outside the path integral.
The generation probability of the inverse move comes from Eq.~\eqref{eq:pgen} after substituting $y$ with $x$.
Since $h_{x,y}\equiv h_{y,x}$, the term inside the path integral and the normalization are the same for both $p_{\mathrm{gen}}(x\rightarrow y)$ and $p_{\mathrm{gen}}(y\rightarrow x)$. The ratio between generation probabilities is
\begin{equation}
    \frac{p_{\mathrm{gen}}(x\rightarrow y)}
{p_{\mathrm{gen}}(y\rightarrow x)}=\frac{b(y)}{\rho(x)}
    \frac{\rho(y)}{b(x)},
\end{equation}
which combined with Eq.~(2) gives
\begin{equation}\label{eq:detailed_balance}
    P_{\mathrm{sp}}(x)~p_{\mathrm{gen}}(x\rightarrow y)
    =P_{\mathrm{sp}}(y)~p_{\mathrm{gen}}(y\rightarrow x).
\end{equation}
Eq.~\eqref{eq:detailed_balance} implies that our SP move already preserves detailed balance. Thus, by accepting all trials, the MCMC sampling scheme converges to $P_{\mathrm{sp}}(x)$ of Eq.~(2) of the Main Text.

\section{Reweighting the Paths}

\subsection{Interface Representation and Correction Factor}

In this section, we provide a justification for Eq.~(6) in Main Text and explain the details of the reweighting algorithm. We reweight the simulated excursions based on the RC. We choose to adopt the same strategy as TIS and go from a SP representation to an interface representation \cite{rogal2010reweighted,van2003novel}. For that, we need to sample interface ensembles $\mathcal P_{\lambda}[\textbf{x}]$. Given a RC value $\lambda$, $\mathcal P_{\lambda}[\textbf{x}]$ is the PE subset of all excursions crossing the isosurface $\lambda$ (i.e., the reactive interface defined by $\lambda$). The sub-subsets $\mathcal P_{\mathrm{A},\lambda}[\textbf{x}]$ and $\mathcal P_{\mathrm{B},\lambda}[\textbf{x}]$ contain only the excursions from A and from B, respectively. By populating the many $\mathcal P_{\lambda}[\textbf{x}]$ all across $\mathrm{R}$, we can properly reweight the paths and efficiently estimate the PE.

TIS and its variants directly sample interface ensembles \cite{van2003novel,cabriolu2017foundations}. However, this is not the case for waste-recycling TPS and RFPS. Despite that, it is still possible to associate each path $\textbf{x}_i$ to an interface ensemble. Following the approach as Ref.~\cite{lazzeri2023molecular}, we introduce a continuous, infinite set of infinitesimal interfaces throughout the reactive space. Then, if $x_i$ is the path's SP, we define the \textit{shooting interface} $\lambda_i=\lambda(x_i)$, and assign $\textbf{x}_i$ to $\mathcal P_{\lambda_i}[\textbf{x}]$. However, a simple association is not enough. To correctly recover the ensemble distribution, we must assign a correction factor $f_i$ to the path.

In RFPS, we enforce a distribution on the SPs, $P_{\mathrm{sp}}(x)$, rather than on the paths. However, the fact that the SPs are Boltzmann-distributed along the shooting interfaces guarantees good properties for the paths and offers a solution for $f_i$. Following Falkner et al.~\cite{falkner2023conditioning}, we derive the (unweighted) path ensemble sampled by RFPS:
\begin{equation}
\begin{aligned}
    \mathcal P_{\mathrm{RFPS}}[\textbf{x}] \propto h_{\mathrm{R}}[\textbf{x}]\sum_{s=1}^{L[\textbf{x}]-1} P_{\mathrm{sp}}(\textbf{x}(s\Delta t))\left[~\prod_{j=s}^{L[\textbf{x}]-1} p_{\Delta t}(\textbf{x}(j\Delta t)\rightarrow\textbf{x}((j+1)\Delta t))~\right.\\
   \times \left.\prod_{j=0}^{s-1}~\overline{p}_{\Delta t}(\textbf{x}((j+1)\Delta t)\rightarrow\textbf{x}(j\Delta t))\right]&,
\end{aligned}
\end{equation}
where the summation over $s$ accounts for all the possible SPs for path $\textbf{x}$.
Again, we repeatedly apply microscopic reversibility:
\begin{equation}
\begin{aligned}
    \mathcal P_{\mathrm{RFPS}}[\textbf{x}] \propto h_{\mathrm{R}}[\textbf{x}]~\sum_{s=1}^{L[\textbf{x}]-1} P_{\mathrm{sp}}(\textbf{x}(s\Delta t))\frac{\rho(\textbf{x}(0))}{\rho(\textbf{x}(s\Delta t))} \prod_{j=0}^{L[\textbf{x}]-1} p_{\Delta t}(\textbf{x}(j\Delta t)\rightarrow\textbf{x}((j+1)\Delta t)).
\end{aligned}
\end{equation}
By injecting Eq.~(2) from the Main Text, we obtain:
\begin{equation}
\begin{aligned}
    \mathcal P_{\mathrm{RFPS}}[\textbf{x}] \propto h_{\mathrm{R}}[\textbf{x}]~\sum_{s=1}^{L[\textbf{x}]-1} b(\textbf{x}(s\Delta t))~\rho(\textbf{x}(0))~\prod_{j=0}^{L[\textbf{x}]-1} p_{\Delta t}(\textbf{x}(j\Delta t)\rightarrow\textbf{x}((j+1)\Delta t)).
\end{aligned}
\end{equation}
We factor out the PE in the reactive space of Eq.~\eqref{eq:PR}, and get
\begin{equation}
\begin{aligned}
    \mathcal P_{\mathrm{RFPS}}[\textbf{x}] \propto \sum_{s=1}^{L[\textbf{x}]-1} b(\textbf{x}(s\Delta t))~\mathcal P_{\mathrm{R}}[\textbf{x}].
\end{aligned}
\end{equation}
We now focus only on the paths with shooting interface $\lambda_i \approx \lambda$. The subset of $\mathcal P_{\mathrm{RFPS}}[\textbf{x}]$ containing only these paths is
\begin{equation}
\begin{aligned}
    \mathcal P_{\mathrm{RFPS},\lambda}[\textbf{x}] \propto \sum_{s=1}^{L[\textbf{x}]-1} b(\textbf{x}(s\Delta t))~\delta(\lambda(\textbf{x}(s\Delta t))-\lambda)~h_\lambda[\textbf{x}]~\mathcal P_{\mathrm{R}}[\textbf{x}]&\\
    \propto \sum_{s=1}^{L[\textbf{x}]-1} \delta(\lambda(\textbf{x}(s\Delta t))-\lambda)~h_\lambda[\textbf{x}]~\mathcal P_{\mathrm{R}}[\textbf{x}]&,
\end{aligned}
\end{equation}
where we could omit the $b(x)$ term because we do not bias along the $\lambda$ isosurface, but only across $\lambda$ values. Then,
\begin{equation}\label{eq:piratio}
\begin{aligned}
    \mathcal P_{\mathrm{RFPS},\lambda}[\textbf{x}] 
    &\propto n_\lambda^\star[\textbf{x}]~h_\lambda[\textbf{x}]~\mathcal P_{\mathrm{R}}[\textbf{x}]\\
    &\propto n_\lambda^\star[\textbf{x}]~\mathcal P_\lambda[\textbf{x}],
\end{aligned}
\end{equation}
where
\begin{equation}
    n_\lambda^\star[\textbf{x}] = \sum_{s=1}^{L[\textbf{x}]-1} \delta(\lambda(\textbf{x}(s\Delta t))-\lambda)
\end{equation}
is the (unnormalized) density of $\textbf{x}$ frames at $\lambda$. Let $n_\lambda[\textbf{x}]$ be density normalized across the reactive region. It means that the paths with SP at $\lambda$ follow the $ n_\lambda[\textbf{x}]~\mathcal P_{\lambda}[\textbf{x}]$ distribution, and when applying the correction factor
\begin{equation}\label{eq:correction2}
    f_i = 1 / n_{\lambda_i}[\textbf{x}_i],
\end{equation}
they follow $\mathcal{P}_{\lambda}[\textbf{x}]$ instead. Remarkably, an analogous factor appears in the virtual exchange TIS move in Ref.~\cite{brotzakis2019approximating}, which uses waste-recycling TPS.

A justification of Eq.~\eqref{eq:piratio} is that the configurations of $\mathcal{P}_{\lambda}[\textbf{x}]$ on interface $\lambda$ must follow the Boltzmann distribution restricted to the interface. Each path in $\mathcal{P}_{\lambda}[\textbf{x}]$ contributes to such distribution with $\approx n_{\lambda}[\textbf{x}]$ frames. When selecting SPs on the interface, paths with larger $n_{\lambda}[\textbf{x}]$ have a higher chance of being reproduced, as more SPs can lead to the same outcome. To correct for this bias, we counterbalance their appearance rate by a prefactor $f=1/n_{\lambda}[\textbf{x}]$ and recover $\mathcal{P}_{\lambda}[\textbf{x}]$.

\begin{figure}
\begin{minipage}{0.33\textwidth}
\includegraphics[width=\textwidth]{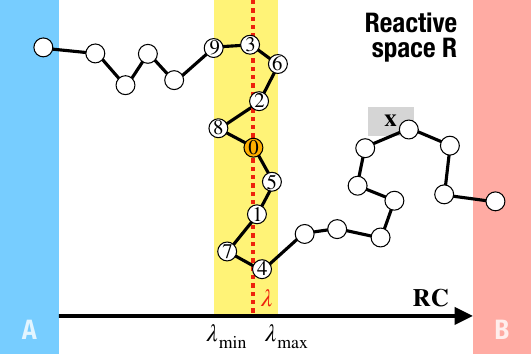}
\end{minipage}
    \caption{$n_{\lambda}[\textbf{x}]$ computation schematics for determining the correction factor $f=1/n_{\lambda}[\textbf{x}]$. The 10 closest frames to $x$ (orange dot) in the RC space are annotated. The alternative SPs at interface ${\lambda}$ that can potentially generate path $\textbf{x}$ are approximately proportional to $n_{\lambda}[\textbf{x}]$.}
    \label{fig:S2}
\end{figure}



The rule for $n_{\lambda}[\textbf{x}]$ must be consistent within the same interface but can change up to a constant across interfaces. One solution involves the inverse speed of interface crossings, as envisioned by Hummer more than two decades ago \cite{hummer2004transition}. In our implementation, given a shooting excursion $\textbf{x}$ we select $n_{\mathrm{neigh}}$ reactive frames with RC value closest to $\lambda$, $n_{\mathrm{neigh}}=\min(10,~L[\textbf{x}]-2)$. Let $\lambda_{\min}$, $\lambda_{\max}$ be the lowest and highest RC value among the selected frames (Fig.~\ref{fig:S2}). Then,
\begin{equation}
    n_{\lambda}^\star[\textbf{x}] = \frac{n_{\mathrm{neigh}}-1}{\lambda_{\max}-\lambda_{\min}}
\end{equation}
is the (unnormalized) density of $\textbf{x}$ at $\lambda$. We aim at uniform average densities across interfaces, as this minimizes the weight difference in the reweighting algorithm and improves the accuracy of the PE estimate. Thus, we finally get
\begin{equation}
    n_{\lambda}[\textbf{x}] = \frac{n_{\lambda}^\star[\textbf{x}]}{\langle n^\star \rangle_\lambda},
\end{equation}
where $\langle n^\star \rangle_\lambda$ is the average unnormalized density for the paths with SP around $\lambda$. The normalization changes as the sampling campaign progresses. If there are more than 10 SPs in the RC range $[\lambda-0.5,\lambda+0.5]$, then $\langle n^\star \rangle_\lambda$ is simply the average of the density of the paths generated from those SPs. Otherwise, the average is computed over the 10 SPs closest to $\lambda$. If $\textbf{x}_i$ comes from a free simulation, $n_{\pm\infty}[\textbf{x}_i]\equiv 1$ and $f_i\equiv 1$, consistently with the fact that all paths from an infinitely long trajectory would have the same weight.

\subsection{Crossing Probability Estimate from Path Sampling with the ``Drop Method''}

\begin{figure}
\begin{minipage}{0.33\textwidth}
\includegraphics[width=\textwidth]{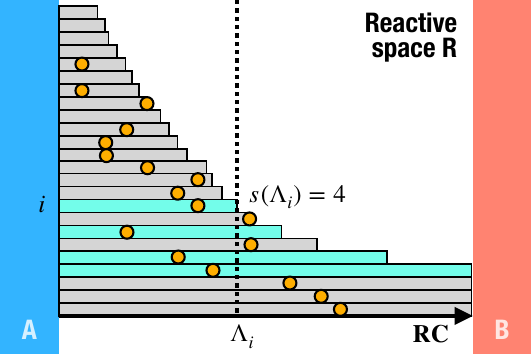}
\end{minipage}
    \caption{$P_{\mathrm{A}}$ computation schematics via the ``drop method''. The sampled segments are rectangles, with SPs as dots if the segments originate from two-way shooting. We focus on the crossing probability drop at $\Lambda_i$: the $s_{\mathrm{A}}(\Lambda_i)=4$ ``survivor'' segments contributing to the crossing probability drop are in cyan, while the others are gray. Thus, the drop is $(s_{\mathrm{A}}(\Lambda_i)-1)/s_{\mathrm{A}}(\Lambda_i)=3/4$.}
    \label{fig:S3}
\end{figure}

The (equilibrium) crossing probabilities $P_{\mathrm{A}}(\lambda)$ and $P_{\mathrm{B}}(\lambda)$ are necessary for reweighting the paths in interface representation. We estimate $P_{\mathrm{A}}(\lambda)$ and $P_{\mathrm{B}}(\lambda)$ by a maximum likelihood procedure on the  sampled excursions. All the excursions from the free simulations around $\mathrm{A}$, the forward shooting snippets with backward shooting simulation reaching $\mathrm{A}$, and the backward shooting snippets with forward shooting simulation reaching $\mathrm{A}$ carry information about $P_{\mathrm{A}}$. For each of those segments (``free'' excursions or half shooting paths), we can obtain the SP values $\lambda_i$ as in Eq.~(5) of the Main Text and extreme values $\Lambda_{i}$ as  
\begin{equation}
    \Lambda_i = \begin{cases}
        \underset{0\leq j\leq L[\textbf{x}_i]}{\max}~\{\lambda(\textbf{x}_i(j\Delta t))\}&\Leftrightarrow~\textbf{x}_i~\text{from}~\mathrm{A},\\
        \underset{0\leq j\leq L[\textbf{x}_i]}{\min}~\{\lambda(\textbf{x}_i(j\Delta t))\}&\Leftrightarrow~\textbf{x}_i~\text{from}~\mathrm{B}.
    \end{cases}
\end{equation}
To factor out the systematic error introduced by the non-infinitesimal time interval between the saved frames, we only consider segments with $\Lambda_i>\lambda_i$.

We do not directly compute $P_{\mathrm{A}}$, but rather its relative drop going through each $\Lambda_{i}$. From the definition of crossing probability, $P_{\mathrm{A}}(-\infty\equiv\mathrm{A})=1$, as all excursions must cross the boundary of $\mathrm{A}$. At each $\Lambda_{i}$, there are $s_{\mathrm{A}}(\Lambda_{i})$ ``survivors'' with SP value $\lambda_{i}<\Lambda_{i}$ that reach a maximum value $\geq \Lambda_{i}$. Past $\Lambda_{i}$, the survivors would be decreased by one unit (Fig.~\ref{fig:S3}). Thus,
\begin{equation}
    P_{\mathrm{A}}(\Lambda_i+\epsilon) = P_{\mathrm{A}}(\Lambda_i)\frac{s_{\mathrm{A}}(\Lambda_i)-1}{s_{\mathrm{A}}(\Lambda_i)}, 
    \quad\forall\Lambda_i < +\infty \equiv \mathrm{B}
    \label{eq:dropA}
\end{equation}
with  $0<\epsilon \ll 1$. The proposed computation is numerically equivalent to solving the WHAM equations  within the limit of one interface for each SP. The advantages are that it is much faster, being a closed solution to the crossing probabilities, and that it allows to directly obtain $P_{\mathrm{A}}$ evaluated at the extremes $\Lambda_i$. An equivalent ``drop equation'' is used for the paths from $\mathrm{B}$: we impose $P_{\mathrm{B}}(+\infty\equiv\mathrm{B})=1$, and
\begin{equation}\label{eq:dropB}
    P_{\mathrm{B}}(\Lambda_i-\epsilon) = P_{\mathrm{B}}(\Lambda_i)\frac{s_{\mathrm{B}}(\Lambda_i)-1}{s_{\mathrm{B}}(\Lambda_i)},
    \quad\forall\Lambda_i > -\infty \equiv \mathrm{A},
\end{equation}
with $ 0<\epsilon \ll 1$, and 
where $s_{\mathrm{B}}(\Lambda_{i})$ is the number of segments shot before $\Lambda_{i}$ that reach a maximum value smaller than $\Lambda_{i}$. To mitigate the errors introduced by suboptimal RC models and further speed up the crossing probability computation, we only use the free excursions when close to the origin state. Specifically, we define a parameter $n_{\mathrm{eq}}$, and exclude from Eqs.~\eqref{eq:dropA},~\eqref{eq:dropB} the shooting snippets whose extremes $\Lambda_i$ are smaller than the $n_{\mathrm{eq}}$-th largest free excursion. In the limit of infinite sampling, only the free simulations would contribute to the crossing probability computation. Thus, we will converge to the equilibrium $P_{\mathrm{A}}(\lambda)$ and $P_{\mathrm{B}}(\lambda)$ by design.

After determining \( P_{\mathrm{A}}(\lambda) \) and \( P_{\mathrm{B}}(\lambda) \), we can derive Eq.~(8) of the Main Text for reweighting using a method similar to that described in Ref.~\cite{lazzeri2023molecular}. In that publication, we also discuss how these equations relate to the WHAM equations and RPE theory \cite{rogal2010reweighted}. A key difference in the current approach is that we now have a reliable representation of \( P_{\mathrm{A}}(\lambda) \) near the metastable states, regardless of the accuracy of the committor model and the effects introduced by the finite time intervals between frames. As a result, we can extend crossing probability-based reweighting to cover the entire reactive space. In this way, we can reconduct the paths extracted from free simulations to the same framework as those from two-way shooting simulations.

\section{Details of the RFPS-AIMMD Algorithm}

In this section, we share the details of the RFPS-AIMMD algorithm and the implementation used in this Letter. 

\subsection{Learning the Committor}

In RFPS-AIMMD and in the original AIMMD algorithm, we learn a model of the committor using a neural network (NN). We will refer to the true committor as $p_{\mathrm{B}}(x)$ and to the committor model as $\tilde p_{\mathrm{B}}(x)$. First, we define a feature representation $y(x)$ of the system that captures essential properties of configurations while simplifying the learning process. A good feature representation incorporates roto-translational invariance and other internal symmetries \cite{jung2023machine}, but it does not have to contain detailed prior information. Furthermore, the NN predicts the logit committor $\lambda$ (see Eq.~14 in the Main Text) rather than directly the committor. Thus, we have:
\begin{equation}
    \tilde p_{\mathrm{B}}(x)=\sigma(\lambda(y(x)))=\frac{1}{1+e^{-\lambda(y(x))}},
\end{equation}
where $\sigma(\cdot)$ is the sigmoid function. Since $\lambda$ ranges from negative to positive infinity, it is easier to learn compared to the committor, which is confined between 0 and 1. Additionally, $\lambda$ does not compress the original space measure when approaching states A and B. It is important to note that $\lambda$ approximates the logit committor only in a subset of the reactive space; this approximation is no longer valid near or inside the states. However, this limitation is not an issue as long as the approximation holds in regions relevant for reweighting simulated paths and estimating the PE. In fact, it allows $\lambda$ to remain meaningful even inside the states. We rely on the NN's generalization ability so that $\lambda$ provides a relevant one-dimensional representation throughout the entire space.

We learn the committor by maximizing the likelihood of the shooting results \cite{peters2006obtaining,jung2017transition}. The training set consists of $(x_i, r_i)$ tuples, where $x_i$ represents the SPs and $r_i = (r_{\mathrm{A},i},~r_{\mathrm{B},i})$ is a vector indicating how many times the path $\mathbf{x}_i$ with SP $x_i$ has reached state $\mathrm{A}$ and $\mathrm{B}$. If $r_i = (2, 0)$, the path is an A-to-A excursion. If $r_i = (0, 2)$, it is a B-to-B excursion. If $r_i = (1, 1)$, it is a transition. In general, $r_{\mathrm{A},i} + r_{\mathrm{B},i} \leq 2$. It can be $r_{\mathrm{A},i} + r_{\mathrm{B},i} < 2$ when the two-way shooting simulation is aborted due to reaching the maximum allowed length. The training loss of Eq.~(13) in the Main Text is the inverse of the log-likelihood for a combined binomial process, which reflects the shooting procedure and the frequentist definition of the committor \cite{peters2010recent,peters2017reaction}:
\begin{equation}\label{eq:loss}
    \mathcal L = -\sum_{i=1}^n r_{\mathrm{A},i}~\log (1-\tilde p_{\mathrm{B}}(x_i)) + r_{\mathrm{B},i}~\log \tilde p_{\mathrm{B}}(x_i).
\end{equation}

To augment the training set, especially close to the metastable states, we include ``virtual'' SPs from the free simulations. Configurations from free simulations around A contribute with $r_i = (2, 0)$, while those around B contribute with $r_i = (0, 2)$. To prevent overfitting due to an imbalance of true and virtual SPs, we ensure equal proportions of each in the training set batches. In each training epoch, we build a batch of size $n_{\mathrm{batch}}$ by randomly selecting true and virtual shooting results with equal probabilities. Training set elements can appear multiple times in the same batch, especially in the early stages of the run. We then minimize the loss for that batch and repeat for a fixed number $E$ of epochs. Unlike previous AIMMD implementations \cite{jung2023machine}, here we reset the NN parameters and retrain from scratch after each committor re-learning step, using all available training data up to that point.

\subsection{Optimizing the SP Selection Bias}
\label{sec:bias}
In RFPS, the selection bias requires to know the equilibrium density projected on $\lambda$, $\rho(\lambda) = e^{-\beta F(\lambda)}$. We estimate $\rho(\lambda)$ by reweighting the paths from our RFPS-AIMMD simulations. The original AIMMD algorithm still controls SP selection but in the context of enhanced TPS. For this purpose, it uses a different bias,
\begin{equation}
    b^\star(x) \propto 1/\rho_{\mathrm{TPE}}(\lambda(x)),
\end{equation}
based on the TPE density projection, $\rho_{\mathrm{TPE}}(\lambda)$, which we evaluate directly on the Markov chain of transition paths \cite{jung2017transition}. In both RFPS-AIMMD and AIMMD, the SP selection bias is the inverse of the target density, whereas acceptance or rejection of a MCMC move depends on the algorithm. With RFPS, all moves are accepted.

In practice, when selecting a SP from a path, we divide the reactive region $\mathrm{R}$ into $n_{\mathrm{bins}}$ equally spaced bins between the limiting isosurfaces $\lambda_{\mathrm{A}}$ and $\lambda_{\mathrm{B}}$, compute the density in each bin, and assign a constant bias to all path configurations within a bin. To accelerate the exploration of all the bins, we modify the selection bias with an additional factor inversely proportional to the number of SPs already selected in the bin. This modification becomes irrelevant as the sampling progresses and the bins get adequately populated, but it helps in the early stages of the simulations to yield better PE estimates. We add a small $\eta=0.1$ to the SP number to prevent division by zero.

While $n_{\mathrm{bins}}$ remains fixed, it is beneficial to adaptively optimize $\lambda_{\mathrm{A}}$ and $\lambda_{\mathrm{B}}$ to improve sampling. Specifically, if the free simulations adequately populate the $\lambda_{\mathrm{A}}$ and $\lambda_{\mathrm{B}}$ isosurfaces, narrowing the SP selection range towards the transition state will produce more rare excursions and transitions, enhancing sampling. Conversely, if free simulations struggle to reach the reactive region, lowering $\lambda_{\mathrm{A}}$ and increasing $\lambda_{\mathrm{B}}$ will select more SPs near the metastable states, helping to bridge free paths with path sampling, although at the cost of fewer transitions.

We now explain how we adaptively optimize $\lambda_{\mathrm{A}}$. At any point during an RFPS-AIMMD or AIMMD run, assume we have obtained $m$ transitions from path sampling simulations and some excursions from free simulations around A. Let $I_0$ be the index of the largest free excursion from A (i.e., the excursion from free simulations around A with the largest extreme value $\Lambda_{I_0}$), $I_1$ be the second-largest, and so on. We set \begin{equation}
    \lambda_{\mathrm{A}} = \Lambda_{I_m}.
\end{equation}
We limit $\lambda_{\mathrm{A}}$ to a range between $-10$ and $-0.5$. The lower boundary is used when there are not enough free excursions, while the upper boundary applies when there are not enough transitions or when the energy barrier is very small. We determine $\lambda_{\mathrm{B}}$ in a similar manner, updating the largest extreme definition and limiting $\lambda_{\mathrm{B}}$ between $0.5$ and $10$.

Since committor learning is more challenging near the states, it is possible for the $\lambda_{\mathrm{A}}$ and $\lambda_{\mathrm{B}}$ isosurfaces to intersect the state boundaries, especially in the early stage of a RFPS-AIMMD run. While this does not pose a theoretical problem, since the selection algorithm ensures that the SPs are always selected from the reactive region $\mathrm{R}$, it can slow down convergence. In fact, it may generate very short excursions that are difficult to decorrelate. This issue is specific to the RFPS algorithm, as the SPs from TPS come from transition paths always spanning the entire RC range. The fact that our implementation prevents $\lambda_{\mathrm{A}}$ and $\lambda_{\mathrm{B}}$ from going below $-10$ or above $+10$ partially addresses the convergence issue.

In RFPS-AIMMD, the RC model evolves over time, which introduces another potential issue specific to the RFPS scheme. After a model update, the SP in the Markov chain may move outside the $[\lambda_{\mathrm{A}}, \lambda_{\mathrm{B}}]$ region. As a result, the associated path may no longer have values within that region, complicating the selection of a new SP. This situation typically happens at the beginning of a run when the model is far from convergence.  To address it, our RFPS-AIMMD implementation selects the path configuration with the lowest absolute RC value. In practice, this is equivalent to adding an exponentially decaying correction $\delta b(x) = e^{-M|\lambda(x)|}$ to the bias, with $M \gg 1$, ensuring that internal consistency is maintained.

\clearpage
\section{Computational Methods}

\subsection{Wolfe-Quapp 2D System}

The Wolfe-Quapp (WQ) potential energy surface (PES, Fig.~2A in the Main Text) has equation:
\begin{equation}\label{eq:wq}
U(x, y) = 2~(x_r^4 + y_r^2 - 2x_r^2 - 3y_r^2 + x_ry_r + 0.3x_r + 0.1y_r),
\end{equation}
where $x_r$ and $y_r$ are the Cartesian coordinates rotated by 45 degrees. The energies are in $k_{\mathrm{B}}T$ units. We evolve a particle with unit mass subject to the WQ PES and overdamped Langevin (diffusive) dynamics \cite{sekimoto1998langevin}, as described by the following equations of motion:
\begin{equation}\label{eq:brownian}
\begin{aligned}
    x(t+dt) &= x(t) - \partial_x U(x(t),y(t))~D~dt+\sqrt{2D~dt}~\zeta_x(t)\\
    y(t+dt) &= y(t) - \partial_y U(x(t),y(t))~D~dt+\sqrt{2D~dt}~\zeta_y(t),
\end{aligned}
\end{equation}
where the random fluctuations $\zeta_x(t)$ and $\zeta_y(t)$ are drawn from a normal distribution, and $Ddt=10^{-5}~[L^2]$. We directly implemented Eqs.~\eqref{eq:brownian} in Python (see the Zenodo repository \cite{prl2025zenodo}). We saved trajectory frames every 100 integration steps ($\Delta t = 100 \cdot dt$) in XTC trajectory files. This choice of $\Delta t$ resulted in approximately 1000 frames for each transition path on average.

The metastable states are circles of radius 0.5:
\begin{align}
    \mathrm{A}=&\{(x, y)\mid \sqrt{(x-x_{\mathrm{A}})^2+ (y-y_{\mathrm{A}})^2}\leq 0.5,\\
    \mathrm{B}=&\{(x, y)\mid \sqrt{(x-x_{\mathrm{B}})^2+ (y-y_{\mathrm{B}})^2}\leq 0.5\},
\end{align}
with the centers $(x_{\mathrm{A}}, y_{\mathrm{A}})=(+0.2143, +1.8746)$ and $(x_{\mathrm{B}}, y_{\mathrm{B}})=(-0.2554, -1.8451)$ in the two lowest local minima.
We estimated the reference transition rates from a $1.73\times 10^{11}~[dt]$-long equilibrium simulation. The simulation produced 921 transition, from which we obtained $k_{\mathrm{AB}}^{\mathrm{ref}}=(1.04 \pm 0.05)\times 10^{-9}~[dt^{-1}]$ and $k_{\mathrm{AB}}^{\mathrm{ref}}=(2.02 \pm 0.09\times 10^{-9})~[dt^{-1}]$. We computed the exact free energies and densities directly from Eq.~\eqref{eq:wq}, given the low dimensionality of the system. We obtained the true committor $p_{\mathrm{B}}(x,y)$ by solving the backward Kolmogorov equation:
\begin{equation}
    \nabla \cdot (\nabla U(x,y)+\nabla)~p_{\mathrm{B}}(x,y)=0
\end{equation}
with $p_{\mathrm{B}}(\mathrm{A})=0$, $p_{\mathrm{B}}(\mathrm{B})=1$ boundary conditions by using the relaxation method \cite{covino_molecular_2019}. With that, we could also estimate the TPE densities as \cite{hummer2004transition}:
\begin{equation}\label{eq:tpe}
    \rho_{\mathrm{TPE}}(x,y) \propto P(\mathrm{TP}|x,y)~\rho(x,y) =2~ p_\mathrm{B}(x,y)~(1-p_\mathrm{B}(x,y))~\rho(x,y).
\end{equation}

From the reference simulations, we also extracted two transitions $\textbf{x}_I^{\mathrm{ref}}$ and $\textbf{x}_{II}^{\mathrm{ref}}$ representative of the main and secondary reactive channels. To classify transitions between these channels, we developed a continuous metric based on the Hausdorff distance \cite{taha2015efficient, chen2011clustering}:
\begin{equation}\label{eq:channels_dist}
    c(\textbf{x}) = \frac{d_\mathrm{H}(\textbf{x},\textbf{x}_I^{\mathrm{ref}})-d_\mathrm{H}(\textbf{x},\textbf{x}_{II}^{\mathrm{ref}})}{d_\mathrm{H}(\textbf{x},\textbf{x}_I^{\mathrm{ref}})+d_\mathrm{H}(\textbf{x},\textbf{x}_{II}^{\mathrm{ref}})}
\end{equation}
for determining if a transition $\textbf{x}$ belongs to either channel.  If $c(\textbf{x}) < -0.5$, we assign $\textbf{x}$ to channel I. If $c(\textbf{x}) > 0.5$, we assign it to channel II. We treat channel switches in the same way as rare event transitions between metastable states but in path space instead of configuration space. In Eq.~\eqref{eq:channels_dist},
\begin{equation}
    d_\mathrm{H}(\textbf{x},\textbf{y})=\min(d'_{\mathrm{H}}(\textbf{x},\textbf{y}), d'_{\mathrm{H}}(\textbf{y},\textbf{x})),
\end{equation}
where $d'_{\mathrm{H}}(\textbf{x},\textbf{y})$ is the directed Hausdorff distance between arrays $\textbf{x}$ and $\textbf{y}$ \cite{taha2015efficient}.

\begin{figure}
\begin{minipage}{\textwidth}
\includegraphics[width=\textwidth]{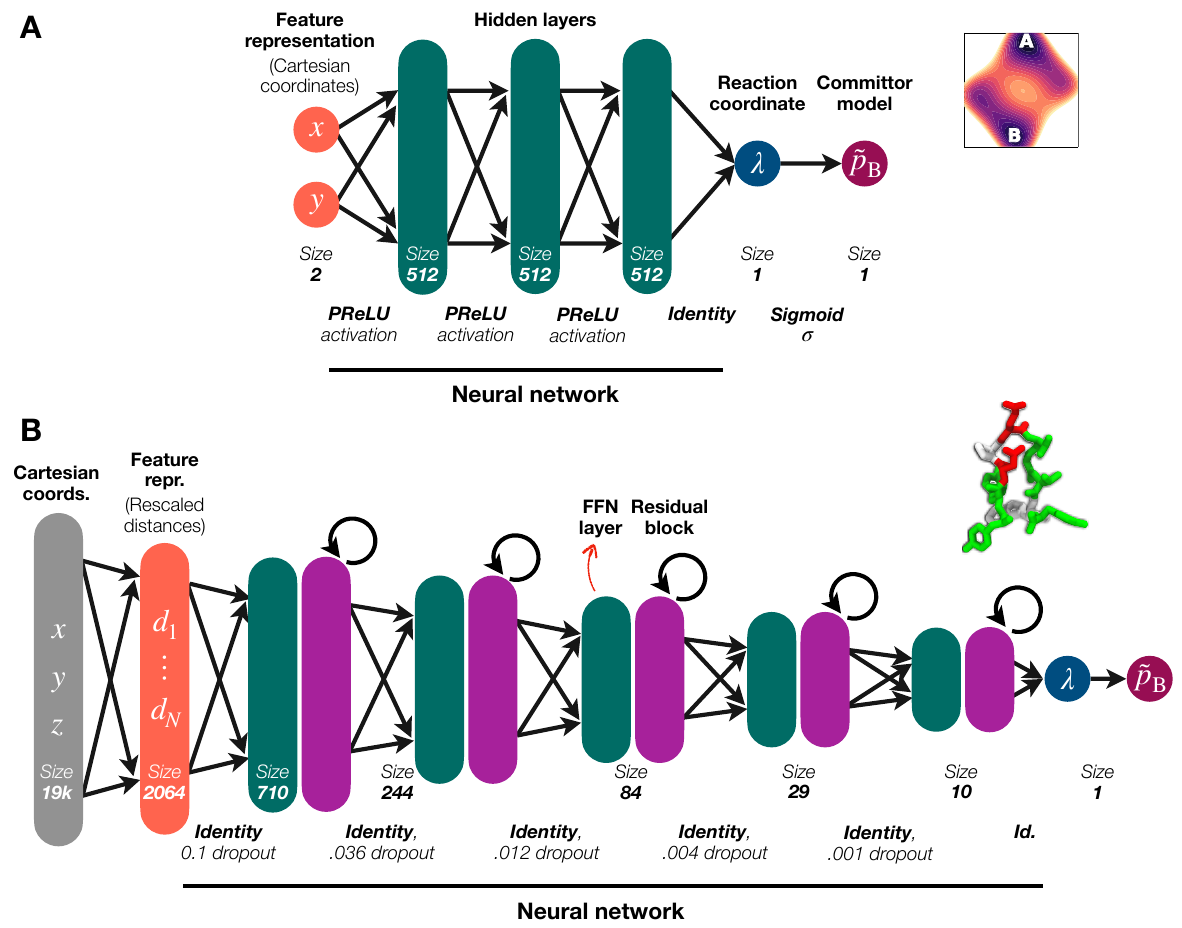}
\end{minipage}
    \caption{\textbf{(A)} WQ system and
    \textbf{(B)} chignolin NN architectures for modeling the committor function.
    }
    \label{fig:S9}
\end{figure}

The initial path for all the path sampling runs is ${\textbf{x}}_{\mathrm{I}}$, then excluded from the PE. In AIMMD or RFPS-AIMMD, the RC model is a feed-forward neural network (FFN) with three hidden layers of 512 neurons and PReLU activation functions \cite{nwankpa2018activation,ding2018activation} (Fig.~\ref{fig:S9}A). The input features to the NN are the $(x,y)$ Cartesian coordinates. During (re)training, we minimized the loss in Eq.~\eqref{eq:loss} over 500 epochs, each with a randomly selected batch of 4096 training set elements, allowing possible repetitions within the batch. We used the Adam optimizer \cite{diederik2014adam} with a learning rate of $l_r = 10^{-4}$. Training the network took a comparable amount of time to the average two-way shooting simulation (a few seconds on our workstation). However, due to the current RFPS-AIMMD and AIMMD implementation, training, and PE estimate were performed asynchronously with respect to the SP selection process (see the \hyperref[sec:implementation]{Section} below) and as such, it did not create bottlenecks in the sampling process.

To determine how quickly the sampled transitions equilibrate in the channels, we calculated the root-mean-squared error (RMSE) of the TPE energy estimates. We focused on the top of the free energy barrier, where the two channels are most distinct. We defined a coarse 2D grid with a width of \(0.25~[L]\), and selected bins with average committor values between 0.4 and 0.6. We numerically computed the true TPE energies in these bins by integrating Eq.~\eqref{eq:tpe} and performing Boltzmann inversion. To enhance the significance of our results, we excluded bins with energies more than \(2.5 ~k_{\mathrm{B}} T\) above the minimum energy. This left us with 13 bins that are representative of both reactive channels, with true energies \(\hat F_1, \ldots, \hat F_{13}\). At different stages of our runs, we computed the TPE free energies \(F_1, \ldots, F_{13}\). We then aligned the true and estimated energies so that \(\sum_i \exp(-\beta F_i) = \sum_i \exp(-\beta \hat F_i) = 1\). Finally, we calculated the RMSE using the following equation:
\begin{equation}
    \mathrm{RMSE}=\sqrt{\langle(\min\{F_i,\hat F_i+2.5~k_{\mathrm{B}}T\}-\hat F_i)^2\rangle},
\end{equation}
where the $\min\{F_i,\hat F_i+2.5~k_{\mathrm{B}}T\}$ caps the error in case the simulated transitions have not adequately populated a bin yet. This cap still allows us to appreciate the difference between fast- and slow-equilibrating runs. In particular, a run that produced transitions only in the main reactive channels would yield \(\mathrm{RMSE} > 1.5~k_{\mathrm{B}} T\).

\clearpage
\subsection{Chignolin}
We followed the protocol of Ref.~\cite{lazzeri2023molecular}. In particular, we took the folded structure of CLN025 (amino acid sequence YYDPETGTWY) from the 2RVD entry of Protein Data Bank \cite{kato2015nmr,yasuda2014physical}. We solvated the peptide with TIP3 water in a $4$ nm cubic periodic box and generated a topology file with Charmm-GUI \cite{sunhwan2008charmmgui}; the final system had 6,468 atoms, 166 belonging to the peptide. We reproduced the settings of Lindorff-Larsen et al. \cite{lindorff2011fastfolding} and chose the CHARMM22$^\star$ force-field \cite{piana2011robust}. We ran the simulations with GROMACS 2022.4 \cite{bauer2022ym} and the velocity Verlet integrator; we fixed the volume after $1$ ns of equilibration and kept the temperature $T=340$ K with the velocity rescale thermostat \cite{bussi2007canonical}. We set a $0.95$ nm threshold for the short-range interactions  and left the remaining GROMACS  parameters unchanged.
We integrated the positions and momenta every $dt=2$ fs and saved the former every $\Delta t=100$ ps in TRR trajectory files. We visualized the simulated trajectories with VMD \cite{humphrey1996vmd}.

We defined the folded ($\mathrm{A}$) and unfolded ($\mathrm{B}$) states based on the fraction of native contacts \cite{best2013native}:
\begin{subequations}\label{boundaries}
\begin{align}
    \mathrm{A} &= \{ x \mid Q(x) \geq 0.99\},\\
    \mathrm{B} &= \{ x \mid Q(x) \leq 0.01\},
\end{align}
\end{subequations}
where the reference configuration ($t=55.1$ ns of the first equilibrium MD simulation) is the centroid of the $\mathrm{C}_\alpha$-RMSD folded state cluster \cite{lindorff2011fastfolding}. Only for the purpose of analysis, we also used a $(d_{3,7},d_{2,8})$ CV representation, where $d_{3,7}$ is the distance between the backbone carbonyl oxygen or the third residue in the peptide chain (D3) and the $\alpha$ carbon of G7, and $d_{2,8}$ is the distance between the $\alpha$ carbon of Y2 and the carbonyl oxygen of T8. We calculated the reference free energy profiles and rate constants from 4 equilibrium MD simulations, totaling $120~$µs. The simulation produced 58 transition, from which we obtained $k_{\mathrm{AB}}^{\mathrm{ref}}=0.28 \pm 0.05 $~µs$^{-1}$ and $k_{\mathrm{AB}}^{\mathrm{ref}}=2.5 \pm 0.5$~µs$^{-1}$.

The initial transition for all the enhanced sampling runs is an unfolding event previously obtained in the work of Ref.~\cite{lazzeri2023molecular}. In general, it is always possible to provide annealing simulations or (energy-minimized) interpolated structures connecting the states as an initial path to diminish the computational burden of sampling a transition at equilibrium. The NN model combines five hidden modules, each consisting of a FFN and a single-block residual neural network layer \cite{jung2023machine, chang2018reversible} (Fig.~\ref{fig:S9}B). The layer sizes decrease progressively from 710 to 10, creating a pyramidal structure, with FFN layers also featuring dropout rates that decrease in intensity. The input features to the neural network are the 2064 protein interatomic distances, including all heavy atom pairs that are at least 4 residues apart. These distances are min-max normalized based on values sampled from short free simulations in $\mathrm{A}$ and $\mathrm{B}$. The exact featurization algorithm is available in the Zenodo repository. For (re)training, we used the same Adam optimizer, learning rate constant ($l_r = 10^{-4}$), and batch size (4096) of the WQ system, but we iterated for only 100 epochs. Training the network required negligible computational resources compared to performing the simulations. Specifically, the training time without GPU acceleration was 20 seconds on our HPC cluster, while the average shooting simulation took half an hour.

To assess the committor learning performance throughout the AIMMD and RFPS-AIMMD runs, we extracted a validation set of 100 SPs from independent runs obtained in Ref.~\cite{lazzeri2023molecular}. These SPs are roughly uniformly distributed in committor space. For each SP $x_i$, we performed 10 two-way shooting simulations and recorded the number of times states A ($r_{\mathrm{A},i}$) and B ($r_{\mathrm{B},i}$) were reached ($r_{\mathrm{A},i}+r_{\mathrm{B},i}=20$). We then estimated the committor using the frequentist definition:
\begin{equation}
    \hat p_{\mathrm{B}}(x_i) = \frac{r_{\mathrm{B},i}}{r_{\mathrm{A,i}}+r_{\mathrm{B,i}}},
\end{equation}
and compared it with $\tilde p_{\mathrm{B}}(x_i)$ from the most recent committor model (Figs.~\ref{fig:S14}D,~\ref{fig:S15}D). Additionally, we developed a WRSME metric. Here, the squared errors in the estimates are rescaled by the expected variance of the underlying binomial process:
\begin{equation}
    \mathrm{WRMSE} = \sqrt{\left\langle \frac{(\tilde p_{\mathrm{B}}(x_i)-\hat p_{\mathrm{B}}(x_i))^2}{\hat p_{\mathrm{B}}(x_i) (1-\hat p_{\mathrm{B}}(x_i))+\epsilon}  \right\rangle},
\end{equation}
where $\epsilon = 0.025$ prevents overflows when the true committor is too low or high to allow accurate frequentist estimates with limited shooting simulations.

\subsection{AIMMD and RFPS-AIMMD Runs}
\label{sec:implementation}

\begin{figure}
    \centering
    \includegraphics[width=0.64\linewidth]{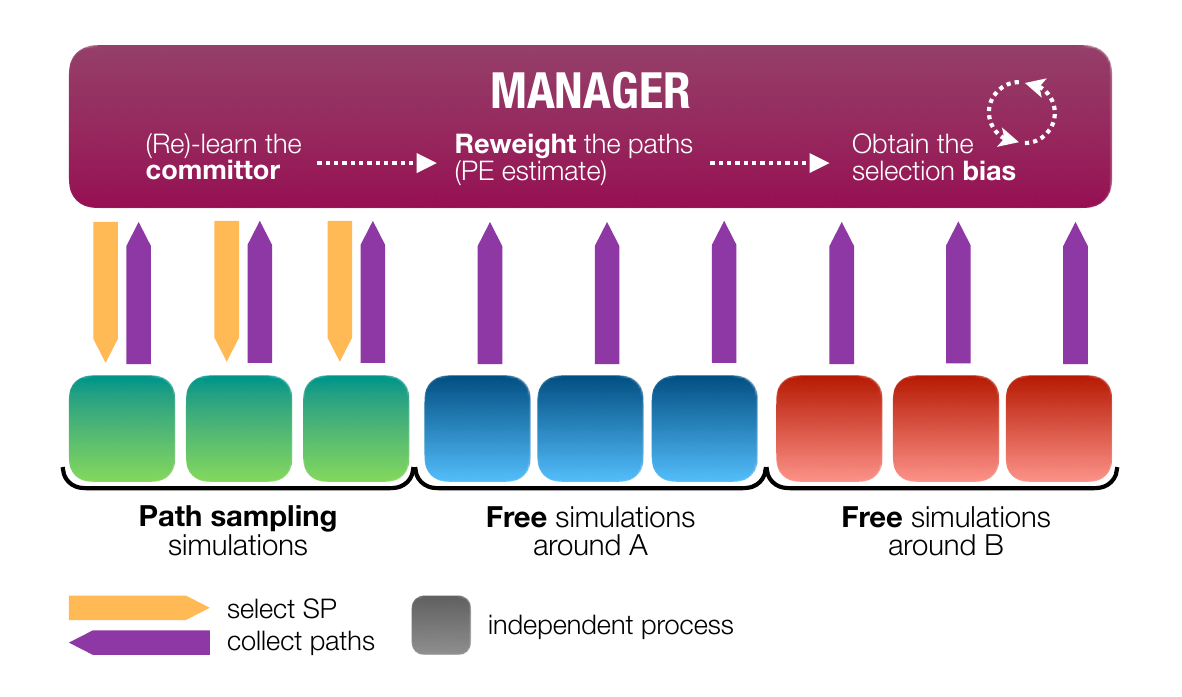}
    \caption{RFPS-AIMMD asynchronous implementation schematics. In this example, three independent processes are dedicated to building separate RFPS Markov chains, three to free simulations around state A, and three to free simulations around state B.}
    \label{fig:manager}
\end{figure}

The code we used for running AIMMD and RFPS-AIMMD enhanced sampling simulations is available in the Zenodo repository \cite{prl2025zenodo}. It is a Python implementation designed for use on both workstations (for 2D systems) and HPC clusters (for MD systems). It uses the MDAnalysis \cite{michaud2011mdanalysis} and MDTraj \cite{mcgibbon2015mdtraj} packages for accessing and analyzing the trajectory files, PyTorch \cite{paszke2019pytorch} for machine learning, for SciPy \cite{virtanen2020scipy} for further data analysis. A custom-written PathEnsemble class collects, processes, and reweights the paths. The repository documentation provides details on how to run the code for the studied systems and how to customize it for new systems and different hyperparameters.

The implementation uses multiple parallel processes (Fig.~\ref{fig:manager}). One set is dedicated to creating path sampling chains and free simulations. An additional ``manager" process handles several tasks: selecting new SPs, collecting the paths, re-learning the committor model from the shooting results, reweighting the paths, and obtaining the selection bias from the committor and reweighted paths. The manager operates asynchronously to ensure efficient execution. SP selection occurs immediately when requested by a path sampling chain, while data collection, committor learning, and reweighting are executed continuously in a loop. This guarantees that the most updated selection bias is always available without delays. This asynchronous structure is especially advantageous for smaller systems, where the time required to generate a new path via two-way shooting is relatively low.

In each enhanced sampling run of this work, we allocated one independent process to path sampling simulations, one to free simulations around state A, and one to free simulations around state B. Each task took roughly one third of the computational budget. At the beginning of the runs, the NN training set had only the boundary frames of the initial transition, and we assumed $\rho(\lambda)\equiv 1$. The initial configurations of the free simulations were taken from the boundary frames of the initial transition of the path sampling chains. Free simulations were interrupted and reinitialized whenever they reached a metastable state different from the target, compatible with how rare are the associated events. During each reinitialization, the process started from the last configuration observed in the target state.

For obtaining the bias, we followed the procedure of the dedicated \hyperref[sec:bias]{Section}. We used $n_{\mathrm{bins}} = 10$ and projected the current PE density (for RFPS-AIMMD) or the TPE density estimate (for AIMMD) onto the RC. In AIMMD-RFPS, we selected the SPs from the latest produced paths, while in AIMMD, we selected them from the latest elements in the TPS Markov chain. The acceptance probability for a $\textbf{x} \rightarrow \textbf{y}$ TPS move is given by:
\begin{equation}
    a(\textbf{x}\rightarrow \textbf{y}) = \min\left\{1,h_{\mathrm{AB}}[\textbf{y}]~\frac{\sum_{j=1}^{L[\textbf{x}]-1}b^*(\textbf{x}(j\Delta t))}{\sum_{j=1}^{L[\textbf{y}]-1}b^*(\textbf{y}(j\Delta t))}  \right\},
\end{equation}
where $b^*(\cdot)$ can change among steps but has to be consistent within a step 
(see Refs.~\cite{jung2017transition, lazzeri2023molecular}). We set the maximum length for a two-way shooting simulation ($10^4$ frames, corresponding to $t_{\max}=10^6~[dt]$ for the WQ system, and $t_{\max} = 1$ µs for chignolin) high enough to ensure that no simulation was prematurely interrupted. In both RFPS-AIMMD and AIMMD, we could reweight the paths and estimate the PE independently of the path sampling algorithm. In the AIMMD method, we waste-recycled the TPS trials for the PE estimates, as in Ref.~\cite{lazzeri2023molecular}. For the crossing probability computation, we used $n_{\mathrm{eq}}=20$ (WQ system) and $n_{\mathrm{eq}}=5$ (chignolin).

\clearpage
\section{Results}

\subsection{Wolfe-Quapp 2D System}

\begin{table}[h]
    \includegraphics[height=.61\textwidth,angle=90,origin=c]{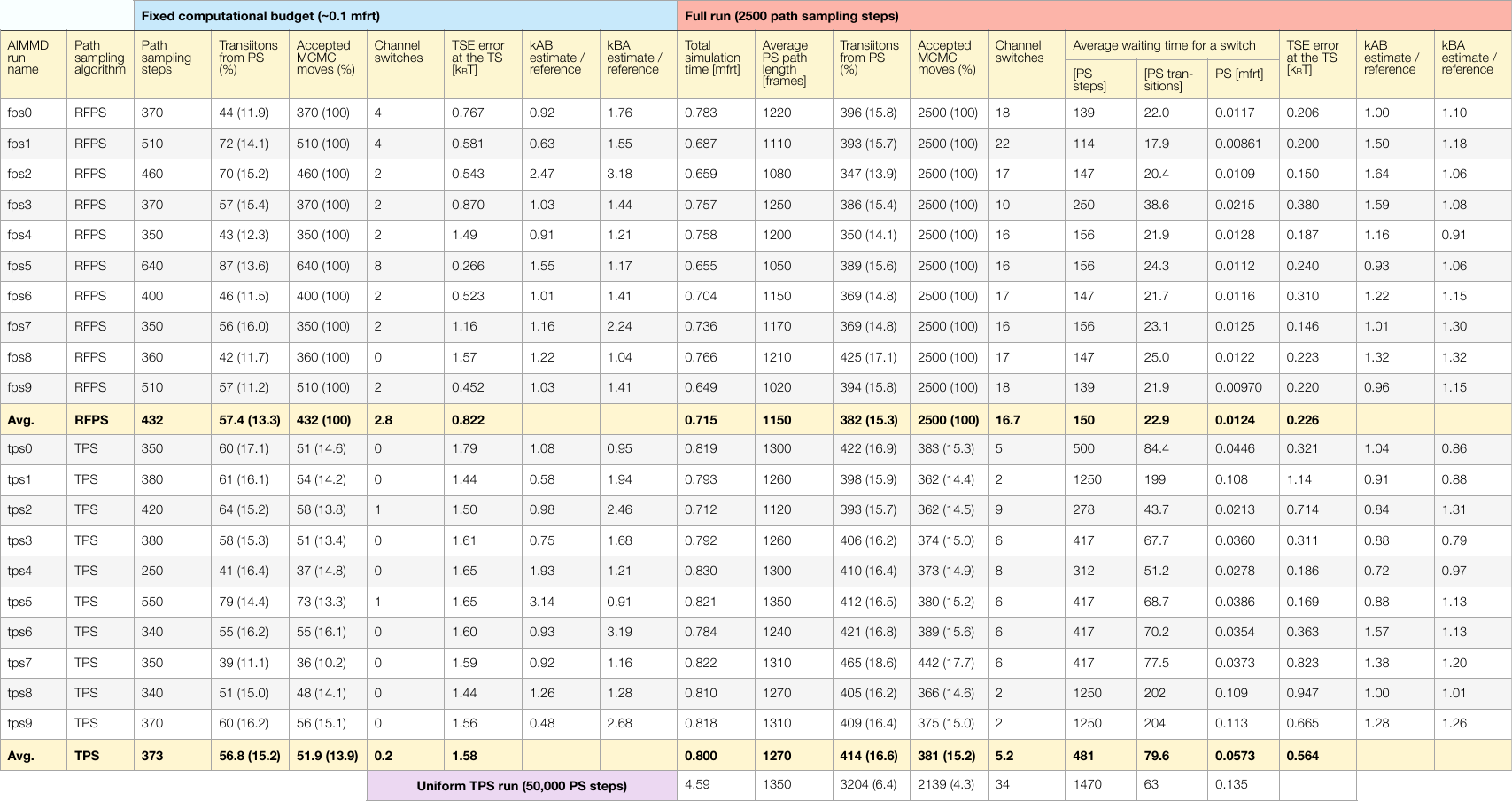}
    \caption{\label{tab:table1}Summary of the enhanced sampling and standard TPS runs on the WQ system. The RFPS-AIMMD runs are \textbf{fps0}, \textbf{fps1}, \dots The AIMMD runs are \textbf{tps0}, \textbf{tps1}, \dots}
\end{table}

\begin{figure}[h]
\vspace{-15pt}
\begin{minipage}{\textwidth}
\includegraphics[width=\textwidth]{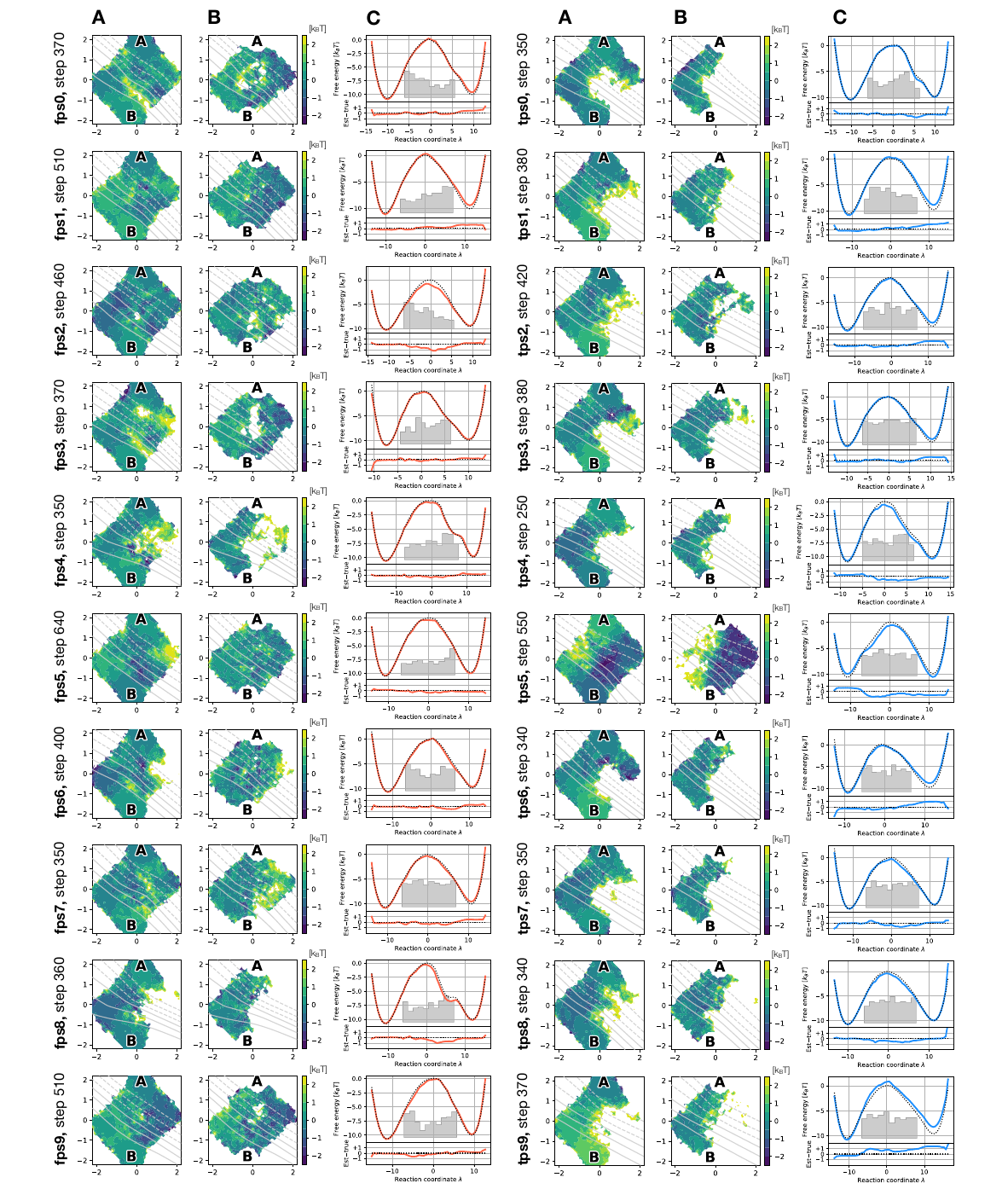}
\end{minipage}
    \caption{Enhanced sampling runs on the WQ system (left: RFPS-AIMMD; right: AIMMD) with fixed budget $\tau \approx 0.1$ mfpt (``data-scarce'' regime).
    The step numbers are annotated in the figure.
    \textbf{(A)} Error on the 2D free energy estimate. The committor model learned at that point of the run is superimposed on the contour plot.
    \textbf{(B)} Error on the 2D TPE free energy estimate.
    \textbf{(C)} 1D free energy estimate projected on the learned committor model (solid line) and true free energy (dotted line). A (linearly rescaled) histogram of the selected SP is superimposed on the plot.}
    \label{fig:S11}
\end{figure}

\begin{figure}[h]
\vspace{-15pt}
\begin{minipage}{\textwidth}
\includegraphics[width=\textwidth]{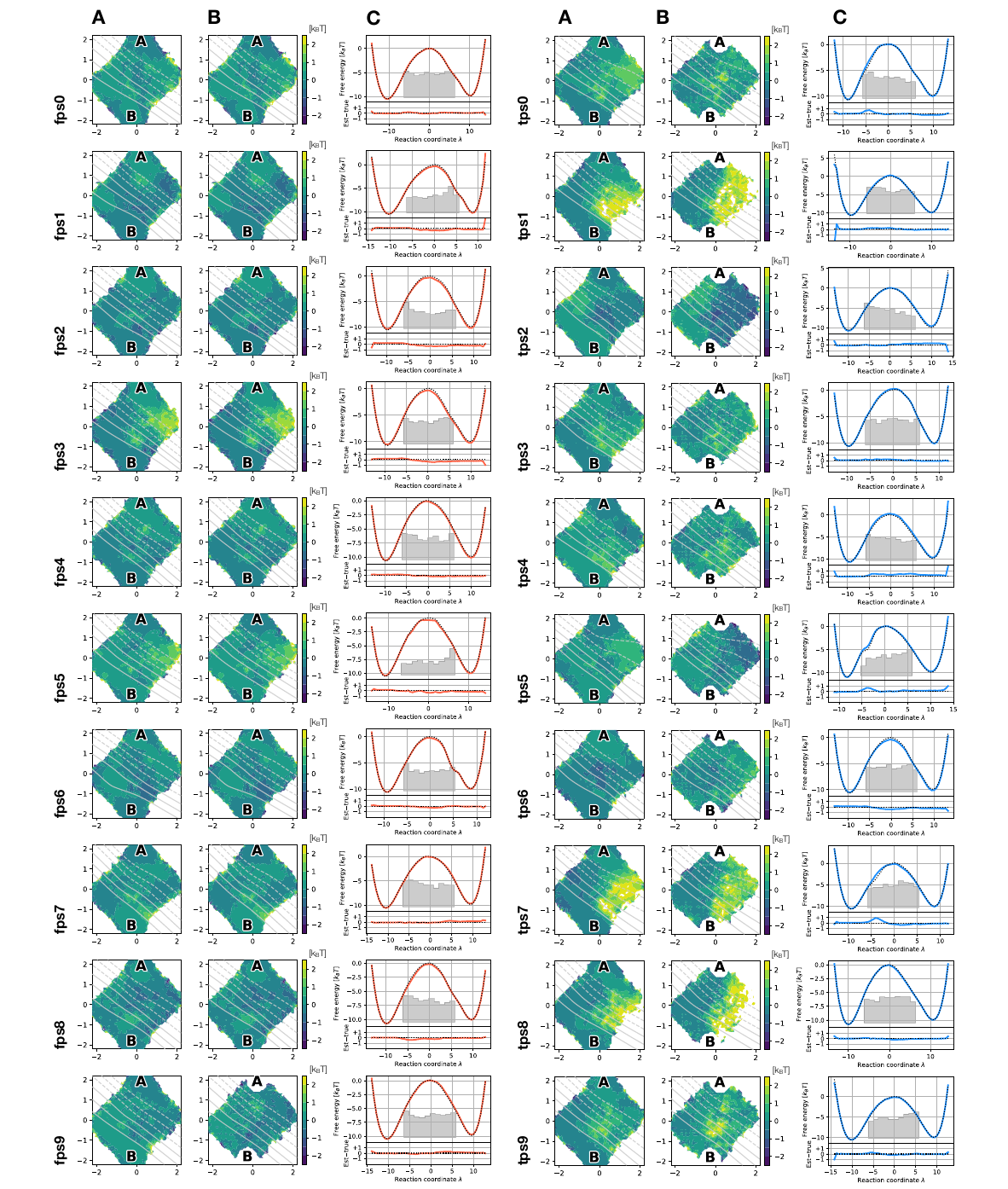}
\end{minipage}
    \caption{Enhanced sampling runs on the WQ system (left: RFPS-AIMMD; right: AIMMD), results after 2500 path sampling steps ($\tau \approx 1$ mfrt).
    \textbf{(A)} Error on the 2D free energy estimate. The committor model learned at that point of the run is superimposed on the contour plot.
    \textbf{(B)} Error on the 2D TPE free energy estimate.
    \textbf{(C)} 1D free energy estimate projected on the learned committor model (solid line) and true free energy (dotted line). A (linearly rescaled) histogram of the selected SP is superimposed on the plot.}
    \label{fig:S12}
\end{figure}


\begin{figure}[h]
\begin{minipage}{\textwidth}
\includegraphics[width=\textwidth]{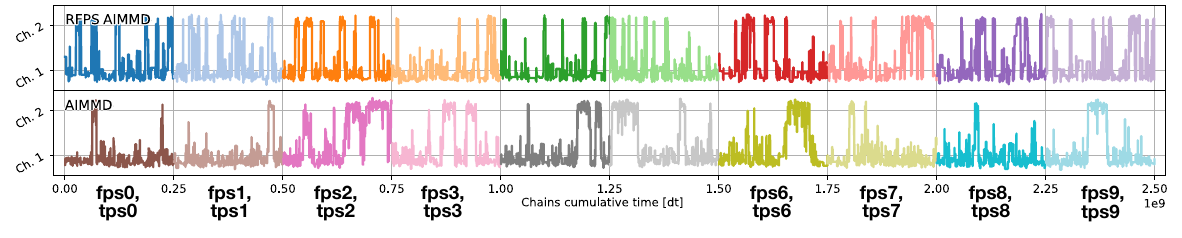}
\end{minipage}
    \caption{Enhanced sampling runs on the WQ system (top: RFPS-AIMMD; bottom: AIMMD). Time series of the reactive channels distance of Eq.~\eqref{eq:channels_dist} for the latest sampled transition. For each run, the results are shown up to $2.5\times 10^8~[dt]$ cumulative time of the path sampling chain, corresponding to approximately 2000 path sampling steps. In general, the average length of the sampled paths varies across runs.}
    \label{fig:S13}
\end{figure}

\clearpage
\subsection{Chignolin}

\begin{table}[h]
    \includegraphics[width=\textwidth]{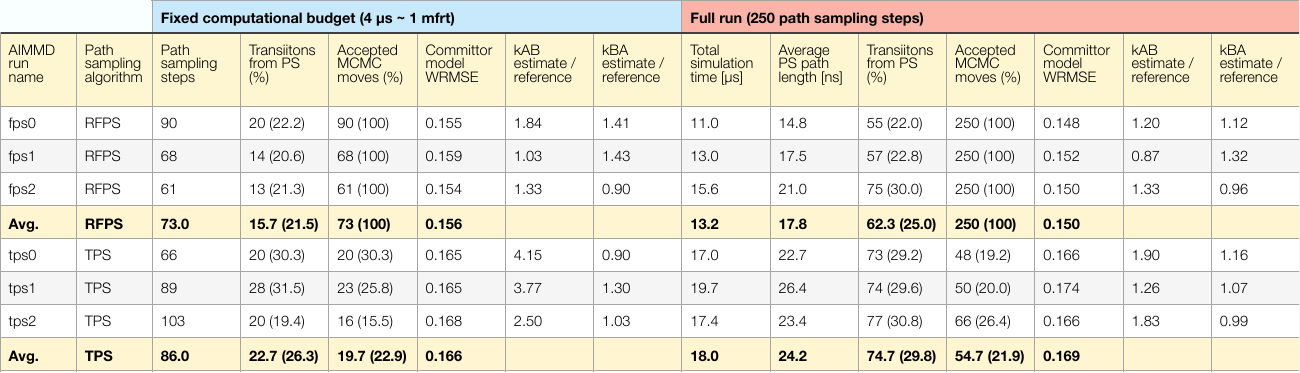}
    \caption{\label{tab:table2}Summary of the enhanced sampling runs on chignolin. The RFPS-AIMMD runs are \textbf{fps0}, \textbf{fps1}, \textbf{fps2}. The AIMMD runs are \textbf{tps0}, \textbf{tps1}, \textbf{tps2}.}
\end{table}

\begin{figure}[h]
\begin{minipage}{\textwidth}
\includegraphics[width=.6\textwidth]{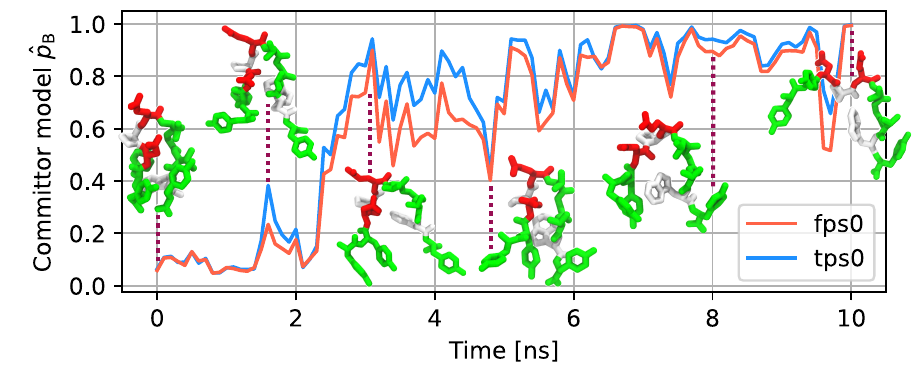}
\end{minipage}
    \caption{Transition from the reference equilibrium simulations projected on two committor models: the last one learned from in RFPS-AIMMD run \textbf{fps0} (red, validation $\mathrm{WRMSE}=0.148$) and the last one learned in the AIMMD run \textbf{tps0} (blue, validation $\mathrm{WRMSE}=0.166$). Representative renders are superimposed on the plot.}
    \label{fig:S16}
\end{figure}

\begin{figure}[h]
\begin{minipage}{\textwidth}
\includegraphics[width=\textwidth]{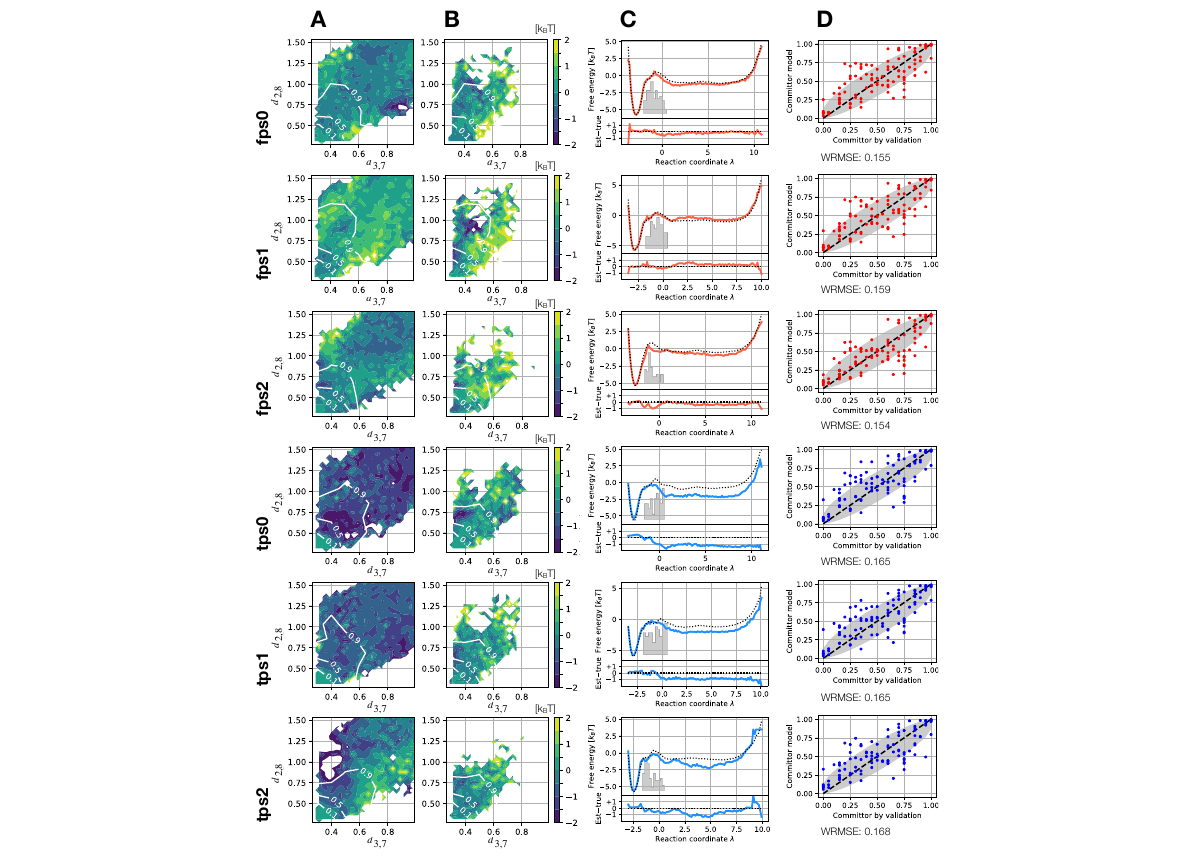}
\end{minipage}
    \caption{Enhanced sampling runs on chignolin (top three: RFPS-AIMMD; bottom three: AIMMD), results with computational budget $\tau = 4$ µs (corresponding to $\approx 1$ mfrt).
    \textbf{(A)} Difference of the free energy estimate projected on the $(d_{3,7},d_{2,8})$ CV set with respect of the equilibrium reference. The committor from PE projections (see Ref.~\cite{bolhuis2011relation}) is superimposed on the plot.
    \textbf{(B)} Difference of the 2D TPE free energy estimate on the $(d_{3,7},d_{2,8})$ CV set with respect of the equilibrium reference.
    \textbf{(C)} 1D free energy estimate projected on the learned committor model (solid line) and reference free energy (dotted line). A (linearly rescaled) histogram of the selected SP is superimposed on the plot.
    \textbf{(D)} Committor model validated on 100 independent SPs. $x$-axis: committor model values. $y$-axis: direct committor computation from 200 shooting events. Gray area: 95\% confidence interval of the direct computation. Based on the discrepancy between the predicted and actual committor values, the WRMSE of the model is annotated on the plot.}
    \label{fig:S14}
\end{figure}

\begin{figure}[h]
\begin{minipage}{\textwidth}
\includegraphics[width=\textwidth]{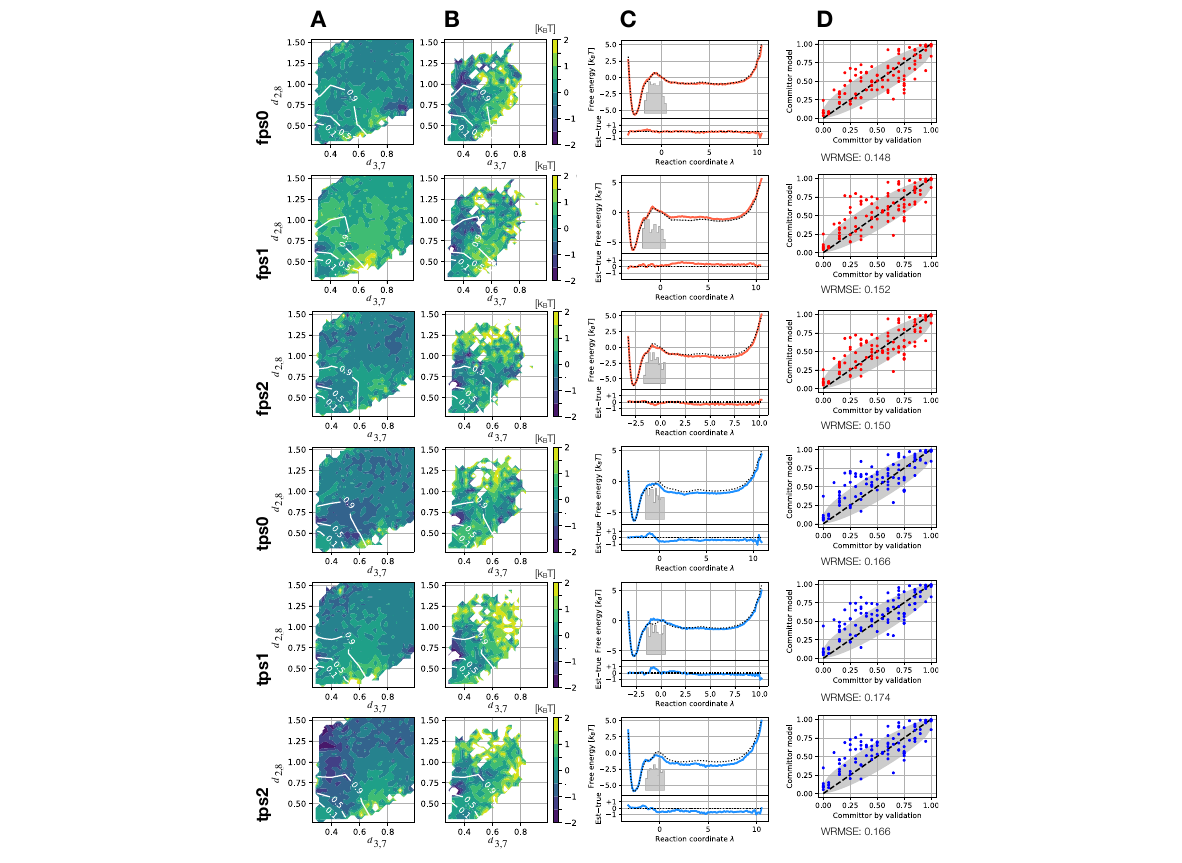}
\end{minipage}
    \caption{Enhanced sampling runs on chignolin (top three: RFPS-AIMMD; bottom three: AIMMD), results after 250 path sampling steps.
    \textbf{(A)} Difference of the free energy estimate projected on the $(d_{3,7},d_{2,8})$ CV set with respect of the equilibrium reference. The committor from PE projections is superimposed on the plot.
    \textbf{(B)} Difference of the 2D TPE free energy estimate on the $(d_{3,7},d_{2,8})$ CV set with respect of the equilibrium reference.
    \textbf{(C)} 1D free energy estimate projected on the learned committor model (solid line) and reference free energy (dotted line). A (linearly rescaled) histogram of the selected SP is superimposed on the plot.
    \textbf{(D)} Committor model validated on 100 independent SPs. $x$-axis: committor model values. $y$-axis: direct committor computation from 200 shooting events. Gray area: 95\% confidence interval of the direct computation. Based on the discrepancy between the predicted and actual committor values, the WRMSE of the model is annotated on the plot.}
    \label{fig:S15}
\end{figure}

\begin{figure}[h]
\begin{minipage}{\textwidth}
\includegraphics[width=.32\textwidth]{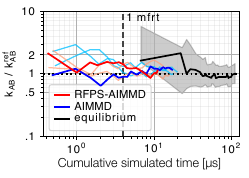}
\end{minipage}
    \caption{Enhanced sampling runs on chignolin. Evolution of the folding rates estimates during the three RFPS-AIMMD (first run: blue; other runs: light blue) and the three AIMMD runs (first run: red; other runs: light red) given the cumulative simulated time. The mfrt scale and results from a long equilibrium run (black) are included for reference. The gray area is the $95\%$ confidence interval of the equilibrium estimate.}
    \label{fig:S21}
\end{figure}


\clearpage
\bibliography{arxiv_250326}